\documentclass[iop,8pt]{emulateapj}
\slugcomment{{\sc Accepted to ApJ:} 2019 January 30}
\usepackage{scrextend} 
\usepackage{float}
\usepackage{natbib}
\usepackage{needspace}
\usepackage{color}

\usepackage{amsmath,amssymb}  
\makeatletter
\newsavebox\myboxA
\newsavebox\myboxB
\newlength\mylenA
\newcommand{\mime}{m_{\rm i}/m_{\rm e}}
\newcommand{\teti}{T_{\rm e0}/T_{\rm i0}}
\newcommand{\betai}{\beta_{\rm i}}
\newcommand{\queirr}{q_{u\rm e, irr}}
\newcommand{\muetot}{M_{u\rm e, tot}}
\newcommand{\muead}{M_{u\rm e, ad}}
\newcommand{\mueirr}{M_{u\rm e, irr}}
\newcommand{\muitot}{M_{u\rm i, tot}}
\newcommand{\muiad}{M_{u\rm i, ad}}
\newcommand{\muiirr}{M_{u\rm i, irr}}

\newcommand{\thetae}{\theta_{\rm e0}}
\newcommand{\thetai}{\theta_{\rm i0}}
\newcommand{\ompinv}{\, \omega_{\rm pe}^{-1}}

\newcommand{\ta}{\, t_{\rm A}}

\newcommand{\Del}{\pmb{\nabla}}
\newcommand{\bigO}{\mathcal{O}}

\newcommand*\xoverline[2][0.75]{%
    \sbox{\myboxA}{$\m@th#2$}%
    \setbox\myboxB\null
    \ht\myboxB=\ht\myboxA%
    \dp\myboxB=\dp\myboxA%
    \wd\myboxB=#1\wd\myboxA
    \sbox\myboxB{$\m@th\overline{\copy\myboxB}$}
    \setlength\mylenA{\the\wd\myboxA}
    \addtolength\mylenA{-\the\wd\myboxB}%
    \ifdim\wd\myboxB<\wd\myboxA%
       \rlap{\hskip 0.5\mylenA\usebox\myboxB}{\usebox\myboxA}%
    \else
        \hskip -0.5\mylenA\rlap{\usebox\myboxA}{\hskip 0.5\mylenA\usebox\myboxB}
    \fi}
\makeatother
\usepackage{csvsimple}
\usepackage{booktabs,multirow,tabularx}
\usepackage{longtable}
\usepackage{color}
\DeclareFontFamily{OT1}{pzc}{}
\DeclareFontShape{OT1}{pzc}{m}{it}{<-> s * [1.10] pzcmi7t}{}
\DeclareMathAlphabet{\mathpzc}{OT1}{pzc}{m}{it}
\usepackage{graphicx}   
\usepackage{bm}
\usepackage{verbatim}   
\usepackage{color}         
\usepackage{subfigure}  
\usepackage{xcolor}
\colorlet{linkequation}{blue}
\usepackage[colorlinks]{hyperref}    
\hypersetup{
	colorlinks = true,
	linkcolor = {blue},
	citecolor = {blue}
	}
	\newcommand*{\SavedEqref}{}
		\let\SavedEqref\eqref
		\renewcommand*{\eqref}[1]{%
  		\begingroup
    		\hypersetup{
      			linkcolor=linkequation,
      			linkbordercolor=linkequation,
    			}%
    		\SavedEqref{#1}%
  	\endgroup
	}

\usepackage{booktabs}
\newcommand{\ra}[1]{\renewcommand{\arraystretch}{#1}}
	
\renewcommand{\b}[1]{\boldsymbol{#1}} 

\newcommand{\unitv}[1]{{\mathbf{\hat{#1}}}}

\def\comp{\,c/\omega_{\rm pe}}
\def\compi{\,c/\omega_{\rm pi}}
\def\nppc{\,N_{\rm ppc}}

\def\tomp{t\omega_{\rm pe}}

\newcommand{\eq}[1]{Eq.~(\ref{eq:#1})}
\newcommand{\fig}[1]{Fig.~\ref{fig:#1}}
\newcommand{\tab}[1]{Tab.~\ref{tab:#1}}
\newcommand{\sect}[1]{Sec.~\ref{sec:#1}}
\newcommand{\app}[1]{App.~\ref{sec:#1}}

\def\bea#1\eea{\begin{align}#1\end{align}}
\newcommand{\makecell}[2][@{}c@{}]{\begin{tabular}{#1}#2\end{tabular}}

\newcommand{\bg}{b_{\rm g}}
\newcommand{\sigmaw}{\sigma_{w}}
\newcommand{\rli}{r_{\rm L i0}}
\newcommand{\rle}{r_{\rm L e0}}
\def\mime{m_{\rm i}/m_{\rm e}}

\def\bvec{\mathbf{B}}
\def\bhat{\mathbf{\hat{b}}}
\def\evec{\mathbf{E}}

\def\sgra{Sgr~A$^*$}
\def\betai{\beta_{\rm i}}
\def\sigmai{\sigma_{\rm i}}
\defcitealias{rsn17}{RSN17}

\begin{document}
\title{Electron and Proton Heating in Transrelativistic Guide Field  Reconnection}
\author{Michael E. Rowan,$^1$ Lorenzo Sironi,$^2$ and Ramesh Narayan$^1$}
\affil{$^1$Harvard-Smithsonian Center for Astrophysics, 
60 Garden Street, Cambridge, MA 02138, USA
\\
$^2$Department of Astronomy, Columbia University, 550 W 120th St, New York, NY 10027, USA}
 \email{E-mail: michael.rowan@cfa.harvard.edu}

\begin{abstract} 
The plasma in low-luminosity accretion flows, such as the one around the black hole at the center of M87 or Sgr A* at our Galactic Center, is expected to be collisioness and two-temperature, with protons hotter than electrons. Here, particle heating is expected to be controlled by magnetic reconnection in the transrelativistic regime $\sigma_{w}\sim 0.1\mbox{--}1$, where the magnetization $\sigma_{w}$ is the ratio of magnetic energy density to plasma enthalpy density. By means of large-scale 2D particle-in-cell simulations, we explore for a fiducial $\sigma_w=0.1$ how the dissipated magnetic energy gets partitioned between electrons and protons, as a function of $\betai$ (the ratio of proton thermal pressure to magnetic pressure) and of the strength of a guide field $B_{\rm g}$ perpendicular to the reversing field $B_0$. At low $\betai\;(\lesssim 0.1)$, we find that the fraction of initial magnetic energy per particle converted into electron irreversible heat is nearly independent of $B_{\rm g}/B_0$, whereas protons get heated much less with increasing $B_{\rm g}/B_0$.
As a result, for large $B_{\rm g} /B_{0}$, electrons receive the overwhelming majority of irreversible particle heating (${\sim}93\%$ for $B_{\rm g} /B_{0}=6$). This is significantly different than the antiparallel case $B_{\rm g}/B_0=0$, in which electron irreversible heating accounts for only ${\sim}18\%$ of the total particle heating \citep[][]{rsn17}. At $\betai \sim 2$, when both species start already relativistically hot (for our fiducial $\sigma_w=0.1$), electrons and protons each receive ${\sim}50\%$ of the irreversible particle heating, regardless of the guide field strength.
Our results provide important insights into the plasma physics of electron and proton heating in hot accretion flows around supermassive black holes. 
\end{abstract}

\maketitle

\section{Introduction} \label{sec:introduction}

When black holes accrete at much below the Eddington limit they tend
to be radiatively inefficient, and the resulting accretion flows
become extremely hot (see \citealt{Yuan2014} for a review). Hot
accretion flows are particularly common in the large population of
low-luminosity active galactic nuclei (\citealt{Ho2008}). Two members of this
population, viz., Sagittarius~A$^*$ (\sgra)---the black hole at the
center of our Galaxy---and the supermassive black hole in M87, are
primary targets of the Event Horizon Telescope (EHT,
\citealt{Doeleman2008, Doeleman2009}), and are of special interest at
the present time.  These systems, and many others like them, can be
modeled within the framework of advection-dominated
accretion flows (ADAFs, \citealt{Narayan1995}; alternatively, radiatively inefficient accretion flows, RIAFs, \citealt{Stone1999,Igumenshchev2003,Beckwidth2008}). However, detailed models, suitable for comparison with observations, require an understanding of electron heating in the accreting plasma, given that the observed emission is powered by electrons; yet, a detailed understanding of electron microphysics is currently lacking.

The key feature of a hot accretion flow is that the accreting gas
heats up close to the virial temperature, causing the flow to puff up
into a geometrically thick configuration and the plasma to become
optically thin.  Because of the low gas density, the plasma is largely
collisionless, i.e., Coulomb collisions between charged particles are
negligible. Furthermore, at radii inside a few hundred $R_{\rm
  S}\equiv 2GM/c^2$, where $M$ is the mass of the black hole and
$R_{\rm S}$ is the Schwarzschild radius, the plasma becomes
two-temperature, with the protons substantially hotter than the
electrons \citep{Yuan2003}.  The two-temperature nature of the gas in
an ADAF is a generic prediction for several reasons: first, electrons
radiate much more efficiently than protons. Second, coupling between
protons and electrons via Coulomb collisions is inefficient at low
densities.  Lastly, compressive heating favors nonrelativistic protons
over relativistic electrons.

Despite these strong reasons, the plasma could still be driven to a single-temperature state if there were additional modes of energy transfer (beyond Coulomb collisions) from protons to electrons.  Several mechanisms of energy transfer in collisionless accretion flows have been proposed, including weak shocks, turbulence, and magnetic reconnection \citep{Quataert1999, Howes2010, Yuan2002, Sironi2015, Sironi2015a, Werner2016, rsn17, Kawazura2018, Zhdankin2018}.  In the present work, we focus on the last of these possibilities, i.e., magnetic reconnection.  

Magnetic reconnection plays an important role in the energy dynamics of numerous astrophysical systems, for example, relativistic jets, hot accretion flows (ADAFs), and coronae above stellar and accretion disk photospheres.  Many of these systems tend to be magnetically dominated, in the sense that $\betai \equiv P_{\rm gas} /P_{\rm mag} \lesssim 1$ (here, $P_{\rm gas}\equiv n_0 k_{\rm B} T_{\rm i0}$ is the thermal pressure of protons, with density $n_0$ and temperature $T_{\rm i0}$, and $P_{\rm mag}\equiv B_0^2/8 \pi$ is the magnetic pressure, with $B_0$ the magnitude of the reconnecting magnetic field).  As a result, the magnetic field is the primary (or at least major) energy reservoir, and energy dissipation may proceed via reconnection.  Although hot accretion flows and disk coronae are magnetically dominated (i.e., low-beta plasmas), the magnetization $\sigma_{w} \equiv 2 P_{\rm mag} / w$ is typically small, with $\sigma_{w} \lesssim 1$; here, $w\equiv(\rho_{\rm e0} + \rho_{\rm i0})c^2 + \hat{\gamma}_{\rm e0} u_{\rm e0} + \hat{\gamma}_{\rm i0} u_{\rm i0}$ is the enthalpy density per unit volume, and $\rho_{\rm e0} = m_{\rm e} n_0, \rho_{\rm i0} = m_{\rm i} n_0, \hat{\gamma}_{\rm e0}, \hat{\gamma}_{\rm i0},$ and $u_{\rm e0}, u_{\rm i0}$ are the rest-mass densities, adiabatic indices, and internal energy densities, respectively, of electrons and protons.  This regime of $\betai \lesssim 1$ and $\sigma_{w} \lesssim 1$, termed \textit{transrelativistic}, provides a unique context for the study of magnetic reconnection, as protons are generally nonrelativistic, whereas electrons can be moderately or even ultra-relativistic \citep{Melzani2014, Werner2016, Ball2018}.  

In a previous work, we explored electron and proton heating in transrelativistic reconnection, for the idealized case of antiparallel fields \citep[][hereafter \citetalias{rsn17}]{rsn17}.  The important question of electron and proton heating has been addressed by others as well, especially in the case of antiparallel reconnection \citep{Melzani2014, Shay2014, Werner2016, Hoshino2018}.  The more general, and astrophysically relevant, case of reconnection includes a guide magnetic field component perpendicular to the plane of the reconnecting field lines.  In fact, recent work suggests that turbulent heating at microscopic dissipation scales may be ultimately  mediated by reconnection \citep{Boldyrev2017, Loureiro2017, Mallet2017, Comisso2018, Shay2018}.  In turbulent systems like accretion flows, turbulent eddies get naturally stretched into thin current sheets, which at small scales become susceptible to the tearing mode instability, which in turn drives reconnection.  At sufficiently small scales, one may then expect that energy dissipation in turbulence is  mediated by reconnection.
At these small scales, the guide field has the same strength as at large scales, yet the strength of the reversing fields becomes smaller at progressively smaller scales in the turbulent cascade. Our work, which focuses on guide field reconnection (up to the regime of strong guide fields), has then broader implications for energy dissipation in a turbulent cascade.

In nonrelativistic reconnection, it has been demonstrated through direct measurements, fully-kinetic and gyrokinetic simulations, and analytical theory, that the strength of the guide field heavily impacts the energy partition between electrons and protons.  In the strong guide field limit, electrons receive a larger fraction of dissipated magnetic energy than protons \citep{Dahlin14, Numata15,Eastwood2018}.  However, for the transrelativistic electron-proton plasma relevant to hot accretion flows and disk coronae, the question is under-explored, yet crucially important for obtaining predictions that can be compared to observations.

To explore the effect of a guide field in transrelativistic reconnection, we use fully-kinetic particle-in-cell (PIC) simulations, which are capable of capturing the fundamental plasma physics that controls electron and proton heating in collisionless systems.  The PIC method captures  from first principles the interplay between charged particles and electromagnetic fields at the basis of reconnection, thereby resolving plasma processes that are out of reach for large-scale magnetohydrodynamic simulations of accretion disks.

In this paper, we investigate the effect of a guide field on electron and proton heating via reconnection in the transrelativistic regime.  We study the dependence of the heating efficiency on the initial plasma properties, by varying the guide field strength and the proton-$\betai$.  For our main runs, we choose $\sigmaw=0.1$ as our fiducial magnetization, and the initial electron-to-proton temperature ratio is set to $\teti=1$.  In a few selected cases, we also vary the temperature ratio ($\teti=0.1$ and $0.3$), as well as the magnetization ($\sigmaw=1$). We employ the realistic mass ratio in all our simulations.

The rest of this paper is organized as follows.
In Section \ref{sec:setup}, we describe the setup and initial conditions of our  simulations.
Next, in Section \ref{sec:methods}, we explain our technique for measuring electron and proton heating at late times, when the energy of bulk motions driven by reconnection has thermalized.
Then, in Section \ref{sec:results}, we summarize the main results of electron and proton heating in guide field reconnection.
We conclude in Section \ref{sec:conclusion}.

\section{Simulation setup} \label{sec:setup}
We use the electromagnetic PIC code \texttt{TRISTAN-MP} \citep{Spitkovsky2005}, which is a parallel version of \texttt{TRISTAN} \citep{Buneman1993}, to perform numerical simulations of magnetic reconnection.  Our simulations are two-dimensional (2D) in space, however all three components of particle momenta and electromagnetic fields are evolved.  In this section, we describe the setup for our simulations of guide field reconnection.  The simulation setup is similar to that described in \citet{Sironi2014}, \citetalias{rsn17}, and \citet{Ball2018} for the study of antiparallel reconnection.  

Simulation coordinates are as follows: the $xy$ plane is the simulation plane; the antiparallel field is along $x$, and the inflow direction is along $y$;  a guide component of magnetic field points in the $z$ direction.

The profile of the antiparallel component of magnetic field is set as $\b{B}_{\rm ap}=-B_0 \tanh(2 \pi y/\Delta_{\rm cs}) \unitv{x}$.  The parameter $\Delta_{\rm cs}$ controls the width over which the antiparallel field $\b{B}_{\rm ap}$ reverses; we usually set $\Delta_{\rm cs}=30 \comp$, where $\comp$ is the electron skin depth,
\bea
	\frac{c}{\omega_{\rm pe}} = \sqrt{\frac{\gamma_{\rm e0}m_{\rm e} c^2}{4 \pi n_0 e^2}}.
\eea
 Here, $\gamma_{\rm e0}m_{\rm e}$ is the electron mass (including relativistic inertia), 
$\gamma_{\rm e0}=1+u_{\rm e0}/(n_0 m_{\rm e} c^2)$, $u_{\rm e0}$ is the initial electron internal energy density, $n_0$ is the number density of electrons (as well as protons) in the inflow, and $e$ is the electron charge.  In all simulations, we use $\comp=4$ (we refer to \citetalias{rsn17} for tests of convergence with respect to the choice of $\comp$).  The magnitude of the antiparallel magnetic field $B_0$ is controlled via the magnetization
\bea \label{eq:sigmaw}
\sigmaw &= \frac{B_0^2}{4 \pi w},
\eea
where 
\bea
w&=n_{0} (m_{\rm e} + m_{\rm i})c^2 + \hat{\gamma}_{\rm e0} u_{\rm e0} + \hat{\gamma}_{\rm i0} u_{\rm i0}
\eea 
is the specific enthalpy density of the inflowing plasma; $\hat{\gamma}_{\rm e0}$, $\hat{\gamma}_{\rm i0}$ are the initial adiabatic indices, and $u_{\rm e0}$, $u_{\rm i0}$ are the initial internal energy densities of electrons and protons.  This definition of magnetization differs (only when plasma is relativistically hot) from the one commonly used in studies of nonrelativistic reconnection, 
\bea
\sigmai&=\frac{B_0^2}{4 \pi n_{\rm 0} m_{\rm i} c^2}.
\eea  
For nonrelativistic temperatures $\sigmaw \cong \sigmai$, however for relativistically hot plasma, \eq{sigmaw} includes the effects of relativistic inertia in the denominator, so that $\sigma_{w} < \sigmai$ in general.  In our simulations, we fix $\sigmaw = 0.1$ (except for a few cases with $\sigmaw=1$, which we explore in \sect{others}).

In addition to the antiparallel field, a guide magnetic field component is initialized perpendicular to the plane of antiparallel field lines, i.e., $\b{B}_{\rm g} = B_{\rm g}\unitv{z}$.  The strength of the guide field is parametrized by the ratio 
\bea
	\bg=B_{\rm g} / B_{0}, 
\eea
where $B_{\rm g}$ is the magnitude of the guide  field (uniform throughout the domain) and $B_0$ is the magnitude of the antiparallel field.
We vary $\bg$ from 0 to 6 (i.e., from the antiparallel case to the strong guide field regime).

Particles in the upstream region are initialized according to a Maxwell-J\"{u}ttner distribution,
\bea
f_{\rm MJ}(\gamma, \theta_{s0}) &\propto \gamma \sqrt{\gamma^2 - 1}\exp(-\gamma/\theta_{s0}),
\eea
where $\theta_{\rm s0}$, $s \in \{ \text{e, i}\}$, is the initial dimensionless temperature of the respective particle species: ${\thetae = k_{\rm B} T_{\rm e0} / m_{\rm e} c^2}$ and ${\thetai = k_{\rm B} T_{\rm i0} / m_{\rm i} c^2}$.  The combination of proton dimensionless temperature $\thetai$ and magnetization $\sigma_{\rm i}$ determines the value of proton-$\betai$,
\bea
\betai &= \frac{8 \pi n_{0} k_{\rm B} T_{\rm i0}}{B_0^2} = \frac{2 \thetai}{\sigmai}.
\eea
Our simulations have $\betai$ in the range $5 \times 10^{-4}$ to 2. For each value, we explore a range of $\bg$ between  and 6.

Magnetic pressure within the current sheet is smaller than that on the outside.  To ensure pressure balance, a population of hot and overdense particles is initialized in the current sheet.  From the pressure equilibrium condition, the temperature of overdense particles in the current sheet, $T_{\rm cs}$, is given by $k_{\rm B} T_{\rm cs} / m_{\rm i} c^2 = \sigmai /2 \eta,$ where $\eta$ is the overdensity of particles in the current sheet, relative to that of the inflowing plasma, $n_0$.  We use $\eta=3$.  Electrons and protons in the current sheet are assumed to have the same temperature.

Parameters associated with each of the main runs are indicated in \tab{params}.  In these simulations, we employ the physical mass ratio, $\mime=1836$, and an initial electron-to-proton temperature ratio in the upstream of $\teti=1$.  In two cases (see \sect{others}), we also consider $\teti=0.1$ and $0.3$, to illustrate the dependence of our results on temperature ratio.

Reconnection is triggered at the center of the box ($x\sim0$, $y\sim0$), by artificially removing the pressure of the hot particles initialized in the current sheet. This leads to the formation of an X-point.  From this central X-point, the tension of reconnected field lines ejects the plasma to the left and to the right, along $x$.  We use periodic boundary conditions along $x$, so the outflows from the two sides (i.e., the moving plasma ejected along $x$ by the field line tension)  meet at the boundary, where their collision forms a large magnetic island (we shall call it ``primary island'' or ``boundary island'').  Here, particles and magnetic flux accumulate, as more plasma reconnects and is ejected along the outflows (this is discussed in detail in \sect{time_evol}).  Along the $y$ boundaries, we use two moving injectors (each receding from $y=0$ at the speed of light) to introduce fresh plasma and magnetic field; the domain is enlarged when the injectors reach the $y$ boundaries.  We refer to \citetalias{rsn17} for further details.

In the present work, we measure electron and proton heating at late times, when the particle internal energies have reached quasi-steady values.\footnote{Thanks to our choice of periodic boundary conditions along $x$, we are able to track the particle energies for extended times, and thus assess the time-asymptotic heating efficiency.}  The primary island is the site where we extract our heating measurements, which requires to run the simulations for a sufficient time such that the outflows from opposite sides of the central X-point meet at the boundary and form the island.
The choice of extracting our heating efficiencies from particles residing in the primary island has advantages and disadvantages. The main disadvantage is that it does not allow to directly probe the heating that results solely from reconnection physics (which was the focus of \citetalias{rsn17}), as it includes, e.g., heating due to shocks generated by the colliding reconnection outflows. On the other hand, our choice is the most appropriate for modeling realistic macroscopic systems, since reconnection outflows are expected to eventually come to rest, and their bulk energy to thermalize (e.g., this is expected in systems like accretion flows, for which the dynamical time and length scales are much larger as compared to that of the reconnection microphysics).
 
We run our simulations up to $t/t_{\rm A} \approx 3\mbox{--}4,$ where $t_{\rm A} = L_{x} / v_{\rm A}$ is the Alfv\'{e}nic crossing time for a box of length $L_{x}$ along the $x$ direction; $v_{\rm A}=c\sqrt{\sigma_{w}/(1+\sigma_{w})}$ is the Alfv\'{e}n speed.\footnote{Note that this definition does not include the effective inertia of the guide field, which could be accounted for by defining an {effective Alfv\'{e}n speed} as $v_{\rm A, eff}=c\sqrt{\sigmaw/[1+\sigmaw(1+\bg^2)]}$.}  In all the cases considered here (even the high guide field cases $\bg \gtrsim 3$, for which the onset of reconnection is delayed due to the large magnetic pressure in the current sheet), we find that evolving the system for around 3--4 Alfv\'{e}nic crossing times is sufficient for the measured temperatures in the primary island to attain quasi-steady values.  The procedure for measuring the heating efficiency is further described in \sect{methods}.  

We find that the time-asymptotic heating efficiencies (especially for protons) are sensitive to the $x$-extent of the domain, if the box is not large enough.  In this case, plasma that is ejected along $x$, away from the center, does not have enough time to reach the expected terminal velocity before stopping at the boundaries. It follows that particle heating in the primary island (which also includes the contribution from thermalization of bulk outflow energy) can be artificially inhibited if the domain is too small.
  We find that a box size $L_{x}\approx 2160 \comp$ is large enough to guard against this effect, and this is the value we use in our simulations; convergence of our heating results with respect to the domain size $L_x$ is discussed in \app{appa}.  In units of the proton skin depth,
\bea
\frac{c}{\omega_{\rm pi}}\! \approx \!\frac{c}{\omega_{\rm pe}} \!\sqrt{\frac{m_{\rm i}}{m_{\rm e}}}\! \left( \!1\!+ \!\frac{\theta_{\rm e0}}{\hat{\gamma}_{\rm e0} - 1}\right)^{-1/2}\!
\left( \!1 \!+ \!\frac{\theta_{\rm i0}}{\hat{\gamma}_{\rm i0} - 1}\right)^{1/2},
\eea
the adopted box size corresponds to at least $L_{x}\approx 51 \compi$, with this lower limit achieved at low $\betai$.  For higher values of $\betai$, the proton skin depth approaches the electron skin depth, and the $x$-extent of the domain approaches $L_{x} \approx 2160 \compi$.  For each value of $\betai$, the box size $L_x$, in units of $\compi$, is listed in \tab{params}.

We use a sufficient number of computational particles per cell $\nppc$ to ensure that numerical heating is negligible with respect to measured heating efficiencies (see \sect{gfheating}; we refer also to \citetalias{rsn17} for convergence tests).  For $\betai$ in the range $5\times10^{-4}$ to $0.5$, we use $\nppc=16$, and for $\betai=2$, we use a larger value, $\nppc=64$.

\begin{table*}
\begin{tabularx}{\linewidth}{c X}
  \toprule \midrule
 \makecell{\textbf{Run ID:} \\ $\betai$ \\ $\bg$ \\ $\theta_{\rm i0}$ \\ $\theta_{\rm e0}$ \\ $\sigma_{\rm i}$ \\ $N_{\rm ppc}$ \\ $\compi$ \\ $\rle$ \\ $\rli$ \\ $L_{x} [c/ \omega_{\rm pi}]$} 
 			& \hfill\makecell{\texttt{b5e-4.bg0} \\ $4.9 \times 10^{-4}$ \\ 0 \\ $2.4 \times 10^{-5}$ \\ 0.045 \\ 0.1  \\ 16 \\ 170 \\ 0.063 \\ 2.6 \\ 51}  
			\hfill\makecell{\texttt{b3e-2.bg0}    \\ 0.031                        \\ 0 \\ 0.0016                      \\ 2.9     \\ 0.1  \\ 16 \\ 58   \\ 0.5     \\ 7.2 \\ 149}
			\hfill\makecell{\texttt{b5e-1.bg0}    \\ 0.5                            \\ 0 \\ 0.031                        \\ 55      \\ 0.12 \\ 16 \\ 14  \\ 2.0     \\ 6.7 \\ 617}
			\hfill\makecell{\texttt{b2.bg0}         \\ 2.0                            \\ 0 \\ 0.39                           \\ 690    \\ 0.36 \\ 64 \\ 5.0 \\ 4.0     \\ 4.9 \\1728}\hfill\null \\     
               \midrule
 \makecell{\textbf{Run ID:} \\ $\bg$ \\ $\rle$ \\ $\rli$} 
 			& \hfill\makecell{\texttt{b5e-4.bg3e-1} \\ 0.3 \\ 0.060 \\ 2.5}  
			\hfill\makecell{\texttt{b3e-2.bg3e-1}     \\ 0.3 \\ 0.48 \\ 6.9} 			
			\hfill\makecell{\texttt{b5e-1.bg3e-1}     \\ 0.3 \\ 1.9 \\ 6.4} 			
			\hfill\makecell{\texttt{b2.bg3e-1}          \\ 0.3 \\  3.8 \\ 4.7} \hfill\null \\     
               \midrule
 \makecell{\textbf{Run ID:} \\ $\bg$ \\ $\rle$ \\ $\rli$} 
 			& \hfill\makecell{\texttt{b5e-4.bg6e-1} \\ 0.6 \\ 0.053 \\ 2.2}  
			\hfill\makecell{\texttt{b3e-2.bg6e-1}     \\ 0.6 \\ 0.43 \\ 6.2} 			
			\hfill\makecell{\texttt{b5e-1.bg6e-1}     \\ 0.6 \\ 1.7 \\ 5.8} 			
			\hfill\makecell{\texttt{b2.bg6e-1}          \\ 0.6 \\  3.4 \\ 4.2} \hfill\null \\     
               \midrule
 \makecell{\textbf{Run ID:} \\ $\bg$ \\ $\rle$ \\ $\rli$} 
 			& \hfill\makecell{\texttt{b5e-4.bg1} \\ 1 \\ 0.045 \\ 1.8}  
			\hfill\makecell{\texttt{b3e-2.bg1}     \\ 1 \\ 0.35  \\ 5.1} 			
			\hfill\makecell{\texttt{b5e-1.bg1}     \\ 1 \\ 1.4    \\ 4.7} 			
			\hfill\makecell{\texttt{b2.bg1}          \\ 1 \\  2.8   \\ 3.5} \hfill\null \\     
               \midrule
 \makecell{\textbf{Run ID:} \\ $\bg$ \\ $\rle$ \\ $\rli$} 
 			& \hfill\makecell{\texttt{b5e-4.bg3} \\  3 \\ 0.02 \\ 0.82}  
			\hfill\makecell{\texttt{b3e-2.bg3}     \\ 3 \\ 0.16 \\ 2.3} 			
			\hfill\makecell{\texttt{b5e-1.bg3}     \\ 3 \\ 0.63 \\ 2.1} 			
			\hfill\makecell{\texttt{b2.bg3}          \\ 3 \\  1.3 \\ 1.6} \hfill\null \\     
               \midrule               
 \makecell{\textbf{Run ID:} \\ $\bg$ \\ $\rle$ \\ $\rli$} 
 			& \hfill\makecell{\texttt{b5e-4.bg6} \\ 6 \\ 0.010 \\ 0.43}  
			\hfill\makecell{\texttt{b3e-2.bg6}     \\ 6 \\ 0.082 \\ 1.2} 			
			\hfill\makecell{\texttt{b5e-1.bg6}     \\ 6 \\ 0.33   \\ 1.1} 			
			\hfill\makecell{\texttt{b2.bg6}          \\ 6 \\  0.66  \\ 0.81} \hfill\null \\     
               \bottomrule
\end{tabularx}
\caption{\raggedright Parameters and values associated with our main simulations, described in \sect{setup}.  The \textit{Run ID} for each simulation is composed of the value of proton-$\betai$ and guide field strength $\bg$.  The electron and proton Larmor radii ($\rle$ and $\rli$) are measured in the upstream.  Parameters listed for antiparallel simulations (those ending in \texttt{bg0}), but not stated for nonzero guide field cases, are implied to be the same. In all cases, $L_{x} = 2160 \comp, \comp=4, \mime=1836, \teti=1$, and $\sigmaw=0.1$.}
\label{tab:params}
\end{table*}

\section{Measurement of late-time heating}\label{sec:methods}
In \sect{time_evol} we discuss the time evolution of the reconnection layer, and in \sect{measure} we discuss the measurement of late-time heating in the primary island.

\subsection{Time Evolution of the Reconnection Layer}\label{sec:time_evol}
\begin{figure*}
		\centering
		\includegraphics[width=\textwidth,clip,trim=0.4cm 0cm 0.4cm 0.0cm]{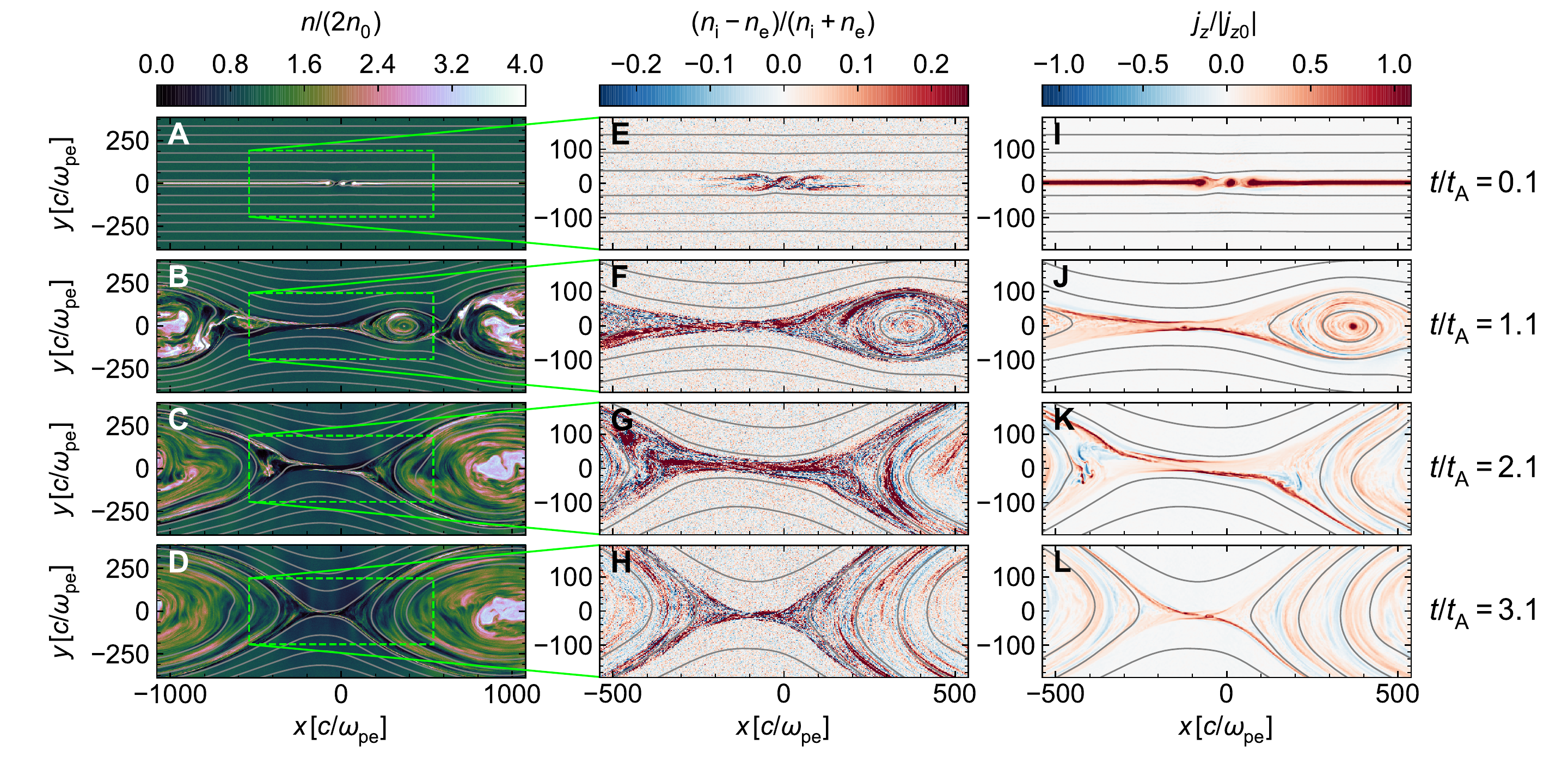} \\
		\caption{Time evolution of a simulation with $\betai=5 \times 10^{-4}$ and $\bg=1$ (\texttt{b5e-4.bg1} in \tab{params}).  The first, second, and third columns show particle number density (in units of initial number density in the upstream), charge non-neutrality, and electric current density along $z$, in units of initial $z$-current in the reconnection layer; the quantities shown here are computed in the simulation frame.  The snapshots are shown at $t/\ta=0.1, 1.1, 2.1$, and $3.1$ (equivalently, $\tomp \approx 7.1\times 10^2, 7.8 \times 10^3 ,1.5 \times 10^4$ and $2.2 \times 10^4$) in the first through fourth rows, as indicated on the right hand side. \label{fig:time_evol2} \\} 
\end{figure*}

In \fig{time_evol2}, we show snapshots covering the time range $t/\ta = 0.1$--$3.1$ for a simulation with $\betai=5\times10^{-4}$ and moderate guide field, $\bg=1$ (run \texttt{b5e-4.bg1} in \tab{params}).  In the first, second, and third columns, respectively, we show the number density $n$ (in units of total upstream density, $2 n_0$), the degree of charge non-neutrality (i.e., the ratio of charge density to particle number density, $(n_{\rm i} - n_{\rm e})/(n_{\rm i} + n_{\rm e})$), and the $z$-component of current density, normalized to the initial value in the current layer, $j_{z}/j_{z0}$; the gray contours show magnetic field lines.

The time evolution is illustrated from the first to the fourth row (e.g., panels A--D).  After reconnection is triggered at $x\sim0$, an X-point forms.  From the central X-point, two reconnection fronts, dragged by magnetic tension, recede from the center; for the simulation shown in \fig{time_evol2}, the speed of recession is $0.31\,c \approx \sqrt{\sigma_{w}}\, c$, so the expected Alfv\'{e}n limit is saturated.  Since we use periodic boundary conditions, the receding fronts meet at $x=\pm 1080 \comp$ after about one Alfv\'{e}nic crossing time (second row), and merge into a volume of particles and magnetic flux that continues to grow as reconnection proceeds.  As anticipated, we refer to this structure as the \textit{primary island}; it is the main site where we extract our heating measurements, since this is where particles ejected from the outflow region eventually end up.  Up to the runtimes of our simulations, the primary island tends to maintain an oblong shape (elongated along $x$), a feature that is more prominent for stronger guide fields. 

Secondary islands, as opposed to the primary island, form frequently at low $\betai$ in the exhaust region (or equivalently, in the outflow region); the formation of secondary islands is suppressed at high $\betai$ (\citealp{Daughton2007}; \citealp{Uzdensky2010}; \citetalias{rsn17}).
 We find that simulations with high guide fields are characterized by a relative absence of secondary islands, as compared to simulations with the same $\betai$ but weaker guide fields.

The current layer in guide field reconnection is characterized by left-right and top-bottom asymmetry, especially in the exhaust region, immediately downstream of the central X-point.  Electrons and protons are ejected from the X-point toward different directions: for our magnetic geometry, electrons to the upper-left and lower-right quadrants, whereas protons are sent to the upper-right and lower-left ones (see panels E--H, which zoom into the central region of panels A--D) \citep{Zenitani2008}.  The $z$-current (third column) is inhomogeneous in the immediate downstream  (see panels I--L); there is some enhancement along the walls of the exhaust (at the interface with the upstream), in particular along the directions that electrons leave the X-point.

\subsection{Measurement of Particle Heating in the Primary Island}\label{sec:measure}
\begin{figure}
		\centering 
		\includegraphics[width=0.45\textwidth,clip,trim=0.5cm 0cm 0.4cm 0.0cm]{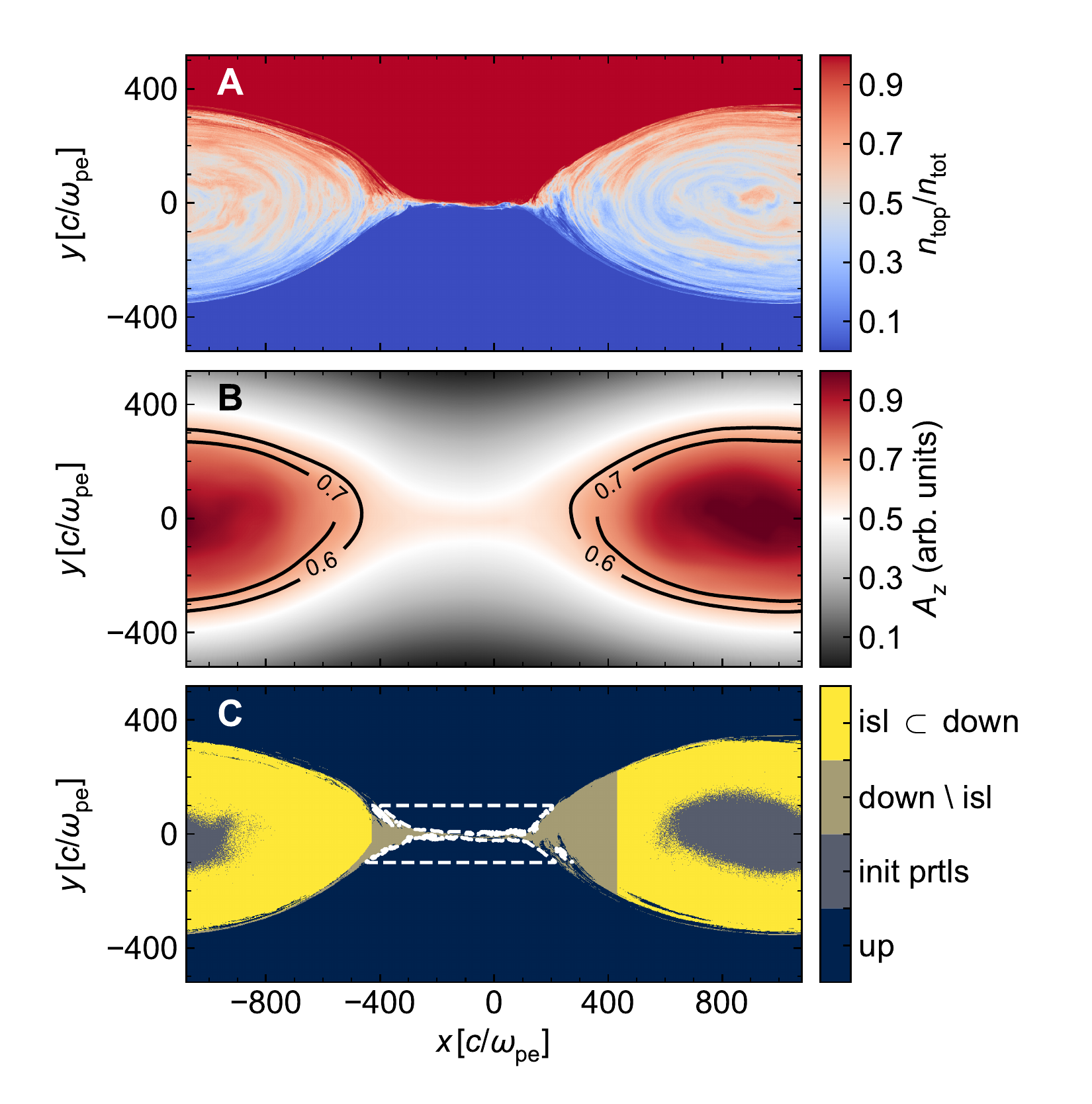} \\
		\caption{2D plots (from simulation \texttt{b5e-4.bg1} in \tab{params}) of (A) ratio of top-to-total particle number density, (B) the $z$-component of magnetic vector potential, and (C) separation into island (yellow), non-island downstream (tan), and  upstream (navy) regions; cells containing particles that are part of the hot overdense population left over from initialization are excluded from the island region, and typically reside at the island core (grey region at the center of the yellow island region in panel C).    In panel B, two contours corresponding to $A_z=0.6$ and 0.7 are plotted (solid black lines) to illustrate the typical shape of magnetic field lines in the primary island (here, units are arbitrary; $A_z$ is normalized to be between 0 and 1). The dashed white contours in panel (C) show the parts of the upstream used to measure inflow quantities. The yellow region that we use for the computation of heating efficiencies in the primary island is defined by criteria (\textit{i}), (\textit{ii}) and (\textit{iii}) described in the text. 
		\label{fig:rec_isl} \\} 
\end{figure}

\begin{figure}
		\centering
		\includegraphics[width=0.45\textwidth,clip,trim=0cm 0cm 0cm 0.0cm]{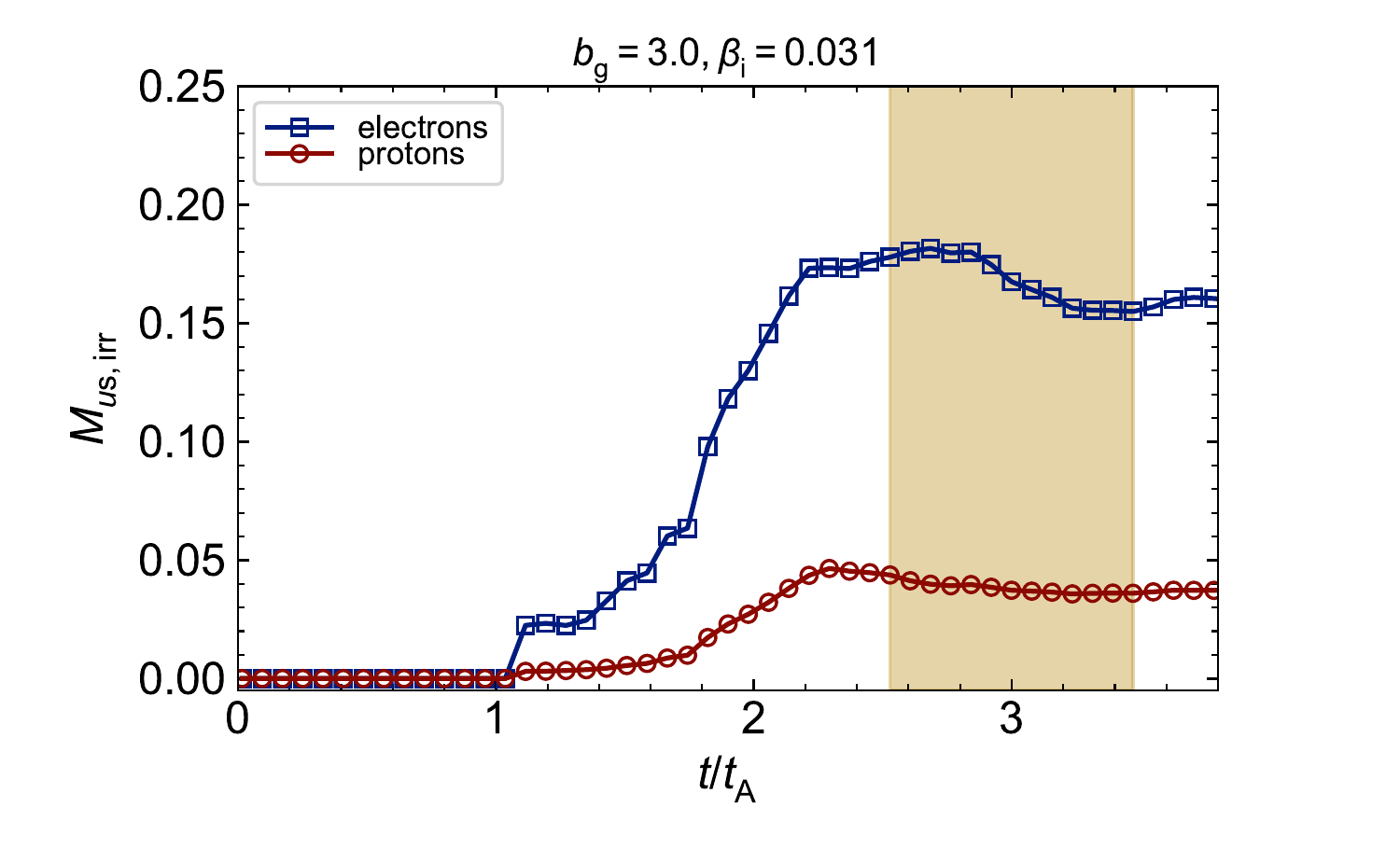} \\
		\caption{Time dependence of irreversible heating fractions $\mueirr$ (blue) and $\muiirr$ (red) for simulation \texttt{b3e-2.bg3}.  The yellow shaded region indicates the time interval used to compute the time-averaged values.\label{fig:mu_bth_timeseries} \\} 
\end{figure}

To assess the heating efficiency at late times, we focus on the change in particle internal energy, as particles travel from the inflow region (i.e., the upstream) to the far downstream, and eventually enter the primary island (these different regions are defined in more detail below). Internal energy and temperature in each cell of the simulation domain are calculated as in \citetalias{rsn17}; here we briefly review the method, but we refer the reader to \citetalias{rsn17} for more details.

The internal energy is computed by treating the plasma as a perfect, isotropic fluid,\footnote{Though we assume isotropy here, our measurements are not significantly affected by this assumption, as we discuss in \sect{anisotropy}.} whose stress-energy tensor is:
\bea \label{eq:pf}
	T^{\mu \nu}_{s} = \left( {e}_{s} + {p}_{s} \right) U^{\mu}_{s} U^{\nu}_{s} - {p}_{s} g^{\mu \nu},
\eea
where $e_{s}{=}\xoverline{n}_{s} m_{s} c^2 + u_{s}, p_{s}, U^{\mu}_{s}$, and $g^{\mu \nu}$ are the rest-frame energy density, pressure, dimensionless four-velocity, and flat-space Minkowski metric, and the subscript ${s}$ denotes the particle species; $\xoverline{n}_{s}$ is the rest-frame particle number density. From \eq{pf}, the dimensionless internal energy per particle in the fluid rest frame ${\upsilon}_{s}$ can be written as
\begin{align} \label{eq:nuint}
	{\upsilon}_{s} &= \frac{(T^{00}_{s}/n_{s} m_{s} c^2 - \Gamma_{s}) \Gamma_{s}}{1 + {\hat{\gamma}}_{s}({\upsilon}_{s}) (\Gamma_{s}^2 - 1)}.
\end{align}
Here,\footnote{To employ Eqs. \ref{eq:pf} and \ref{eq:nuint}, one must choose which frame to boost to; the rest-frame stress-energy tensor is computed from the lab-frame one via Lorentz transformations, 
$T^{\alpha' \beta'} = \Lambda^{\alpha'}_{\;\mu}(\mathbf{v}_{\rm fl}) \Lambda^{\beta'}_{\;\nu}(\mathbf{v}_{\rm fl})T^{\mu \nu}$, where $\mathbf{v}_{\rm fl}$ 
is the local fluid velocity.  This does not necessarily ensure that $T^{0'i'}=T^{i'0'}=0$, so we have tested a more precise, also more expensive, calculation, i.e., solving for $\mathbf{v}$ from $T^{\alpha' \beta'} = \Lambda^{\alpha'}_{\;\mu}(\mathbf{v}) \Lambda^{\beta'}_{\;\nu}(\mathbf{v})T^{\mu \nu}$, subject to the constraints $T^{0'i'}=T^{i'0'}=0$ (by  symmetry of the stress-energy tensor, these are three equations).  The solution of these equations yields a boost $\Lambda(\mathbf{v})$ which ensures $T^{0'i'}=0$.  However, whether we boost to the frame defined by $\mathbf{v}_{\rm fl}$ or $\mathbf{v}$, our results are unchanged.}  $T^{00}_{s}$ is the lab-frame energy density, $n_{s}$ is the lab-frame particle number density, $\Gamma_{s}$ is the Lorentz factor computed from the local fluid velocity, ${\hat{\gamma}}_{s}$ is the adiabatic index, and the subscript $s \in \{ \text{e, i}\}$ indicates the type of particle (electron or proton).  Note that the adiabatic index ${\hat{\gamma}}_{s}({\upsilon}_{s})$ is a function of the specific internal energy.  Given a mapping between the specific internal energy and adiabatic index, \eq{nuint} can be solved iteratively for ${\upsilon}_{s}$.  For adiabatic index, we use a function of the form
\bea
	{\hat{\gamma}}_{s}({\upsilon}_{s}) &= \frac{A+B{\upsilon}_{s}}{C+D{\upsilon}_{s}}, \label{eq:adindex}
\eea
where $A \approx 1.176, B \approx 1.258, C \approx 0.706$, and $D \approx 0.942$.  The numerical coefficients satisfy $A/C = 5/3$ and $B/D = 4/3$ in the nonrelativistic (${\upsilon}_{s}\rightarrow 0$) and ultra-relativistic (${\upsilon}_{s}\rightarrow \infty$) limits, respectively; see Eq. 14 of \citetalias{rsn17} for additional details.  The adiabatic index in \eq{adindex} is used to convert between specific dimensionless internal energy and dimensionless temperature: $\theta_{s} = [{\hat{\gamma}}_{s}(\upsilon_{s}) - 1]\upsilon_{s}$.

In the following, we refer to ``downstream'' as the combination of the outflow region and the primary island. 
We select only part of the downstream to compute the late-time particle heating, in particular part of the primary island, which is far from the central X-point.  The region is selected based on three criteria: (\textit{i}) the mixing between particles originating from the top ($y>0$) and bottom ($y<0$) of the domain must exceed a chosen threshold (and be less than the complementary threshold): $d_{\rm th}< n_{\rm top}/n_{\rm tot} < 1-d_{\rm th}$ \citepalias{rsn17}, (\textit{ii}) the $z$-component of the magnetic vector potential must exceed a value, $A_{z} > A_{z, {\rm th}}$, which is related to the mixing threshold identified in (\textit{i})  \citep{Li2017, Ball2018}, and (\textit{iii}) cells containing particles that were part of the hot, overdense population initialized in the current sheet (see \sect{setup}) are excluded, since their properties depend on arbitrary choices at initialization.  
The use of the above criteria for selection of the ``island'' region is illustrated in \fig{rec_isl}.  
Panel A shows the ratio of density of particles originating from the top of the domain, $n_{\rm top}$, to the total density $n_{\rm tot}$.  The part of the downstream that has mixed, according to (\textit{i}) above, is shown in panel C by the combination of grey, yellow and tan regions.

To select the island area, which is a subset of the mixed region, we find cells at the boundary $x=\pm1080 \comp$ that satisfy $d_{\rm th}< n_{\rm top}/n_{\rm tot} < 1-d_{\rm th}$, and of those cells, we select the ones at the upper and lower edges of the island (along $\pm y$).  In these cells, we compute the average value of the vector potential $A_{z, {\rm th}}$, to serve as a second threshold for selection of the island region (panel B).  The island cells are then identified as those where there is sufficient mixing (criterion (\textit{i})), and where $A_z>A_{z, {\rm th}}$  (criterion (\textit{ii})).  We also impose a strict spatial cutoff on the island region, to ensure that it is distinct from the exhaust even at late times (see \citetalias{rsn17} for details). This criterion corresponds in panel C of \fig{rec_isl} to excluding the tan regions at $|x| <  430 \comp$.  Finally, from the island region, we exclude any cells where the density of the hot, overdense particles used for initialization (see \sect{setup}) is greater than zero (criterion (\textit{iii})), so that the measured heating does not depend on particles whose properties are set by hand as initial conditions.  These initial particles generally reside in the island center (see the grey core in panel C). The region that satisfies all our criteria (which we shall call ``island region'' for brevity)  is shown in panel C of \fig{rec_isl} as the yellow area.


The method of island selection outlined here is a robust and consistent way of selecting cells that are far downstream of the central X-point, for all guide field strengths we consider, and it is relatively insensitive to the choice of threshold value $d_{\rm th}$.  For example, $A_{z, {\rm th}}$ differs by no more than  $10\%$ for $d_{\rm th}$ in the range $0.003$--$0.3$;  the overall measured values of particle heating in the island show comparable sensitivity to the choice of $d_{\rm th}$, at a level of around $15\%$ for $d_{\rm th}$ in the range $0.003$--$0.3$.  For island selection, we find that $d_{\rm th}=0.3$ is suitable.

To assess particle heating, we measure the change in particle internal energies, as they travel from the inflow to the island region (described above).  The upstream region is defined such that $n_{\rm top}/n_{\rm tot}<d_{\rm th, up}$ or $n_{\rm top}/n_{\rm tot}>1-d_{\rm th, up}$ (so, a complementary definition to the mixing criterion (\textit{i}) above). We employ  a threshold value $d_{\rm th, up}=3\times10^{-5}$; the fact that $d_{\rm th, up}<d_{\rm th}$
provides a thin (${\sim}$few $\comp$) buffer region between the downstream (tan and yellow areas combined in panel C of \fig{rec_isl}) and the upstream (navy region in panel C of \fig{rec_isl}). We further select only those upstream cells lying within $\pm100\comp$ of $y=0$ (as delimited by the dashed white contours in panel C of \fig{rec_isl}).  

With the inflow and island regions suitably identified, overall heating fractions can be computed as the difference between the dimensionless internal energies in the island and inflow regions, normalized to the inflowing magnetic energy per particle (\citealp{Shay2014}; \citetalias{rsn17}):
\bea
M_{u\rm e,tot} &\equiv \frac{\upsilon_{\rm e,isl}-\upsilon_{\rm e,up}}{\sigma_{\rm i} m_{\rm i} / m_{\rm e}}, \label{eq:mue} \\
M_{u\rm i,tot} &\equiv \frac{\upsilon_{\rm i,isl}-\upsilon_{\rm i,up}}{\sigma_{\rm i}} \label{eq:mui}.
\eea
These dimensionless ratios indicate the fraction of magnetic energy per particle in the inflow that is converted to particle heating, by the time the particle reaches the island, far downstream of the central X-point.  As in \citetalias{rsn17}, the heating fractions in Eqs. \ref{eq:mue} and \ref{eq:mui} can be decomposed into adiabatic-compressive and irreversible components,
\bea
M_{u\rm e,tot} &= M_{u\rm e,ad} + M_{u\rm e,irr} \label{eq:muedecomp},\\
M_{u\rm i,tot}  &= M_{u\rm i,ad} + M_{u\rm i,irr}. \label{eq:muidecomp} 
\eea
The adiabatic heating fractions represent the heating that results solely from an increase in internal energy due to adiabatic compression of the plasma as it travels from the inflow to the island; 
for electrons, the adiabatic heating fraction is approximately\footnote{This is an approximation because \eq{compapprox} assumes a constant adiabatic index $\hat{\gamma}_{\rm e}$; in reality, when calculating $\muead$, we properly account for the possibility of a changing adiabatic index, as is appropriate for electrons that start nonrelativistic in the upstream, but are heated to ultra-relativistic temperatures by the time they reach the island.}  \citepalias{rsn17}
\bea \label{eq:compapprox}
M_{u\rm e,ad} &\approx \frac{1}{2}\betai \frac{T_{\rm e0}}{T_{\rm i0}}\left[ \left(\frac{n_{\rm isl}}{n_{0}} \right)^{\hat{\gamma}_{\rm e}-1} - 1\right].
\eea
where $n_{\rm isl}$ is the typical electron density in the island.
The irreversible heating fractions are associated with a genuine increase in the entropy of the particles, and are of primary interest to us.  The measured heating fractions we present in \sect{results} are typically time-averaged over one Alfv\'{e}nic crossing time ($\approx 7100\ompinv$).

A representative temporal evolution of electron and proton irreversible heating fractions, $\mueirr$ and $\muiirr$, is shown in \fig{mu_bth_timeseries}.  The time evolution of the heating fractions is shown from $t/\ta=0$ to $t/\ta\approx 3.5$; at late times, the heating fractions achieve a steady state (i.e., both the electron and proton irreversible heating fractions are relatively flat after $t/\ta\approx 2.5$).  Time-averaged heating fractions are computed during this steady state; the points used for time-averaging are indicated by the shaded region in \fig{mu_bth_timeseries}.

\section{Results} \label{sec:results}
In this section, we discuss our measurements of electron and proton heating in the primary island, and their dependence on guide field strength $\bg$ and upstream proton-$\betai$.  In \sect{compare}, we focus on one low and  one high $\betai$ case, and explore the effect of guide field strengths $\bg$ in the range 0.3--6.  Next, in \sect{recrate}, we show the dependence of the reconnection rate on $\betai$ and the guide field.  In \sect{gfheating}, we present comprehensive results of electron and proton heating, extracted from a suite of simulations that span the whole parameter space $\bg=$ 0--6 and $\betai= 5 \times 10^{-4}$--$2$.  Here, we focus on the case of equal initial electron and proton temperatures in the upstream, $\teti=1$.  For these simulations, the magnetization is $\sigma_{w}=0.1$.  
In \sect{others}, we present several results of irreversible electron heating from simulations with temperature ratios in the range $\teti = 0.1$--$1$, as well as several cases with $\sigmaw=1$.
Next, in \sect{fitfunc}, we a provide a fitting function for the electron irreversible heating efficiency, based on the simulation results presented in Secs. \ref{sec:gfheating} and \ref{sec:others}.
Then, in \sect{anisotropy}, we discuss the degree of anisotropy in the particle distribution (as a function of $b_{\rm g}$ and $\beta_{\rm i}$), and its effect on the accuracy of our results.
Lastly, in \sect{mechanism}, we discuss an application of the guiding-center formalism to dissect the mechanisms responsible for electron heating at low $\betai$.

\subsection{Electron and Proton Heating: Weak vs. Strong Guide Field}\label{sec:compare}
\begin{figure*}
		\centering
		\includegraphics[width=\textwidth,clip,trim=0cm 0cm 0cm 0.0cm]{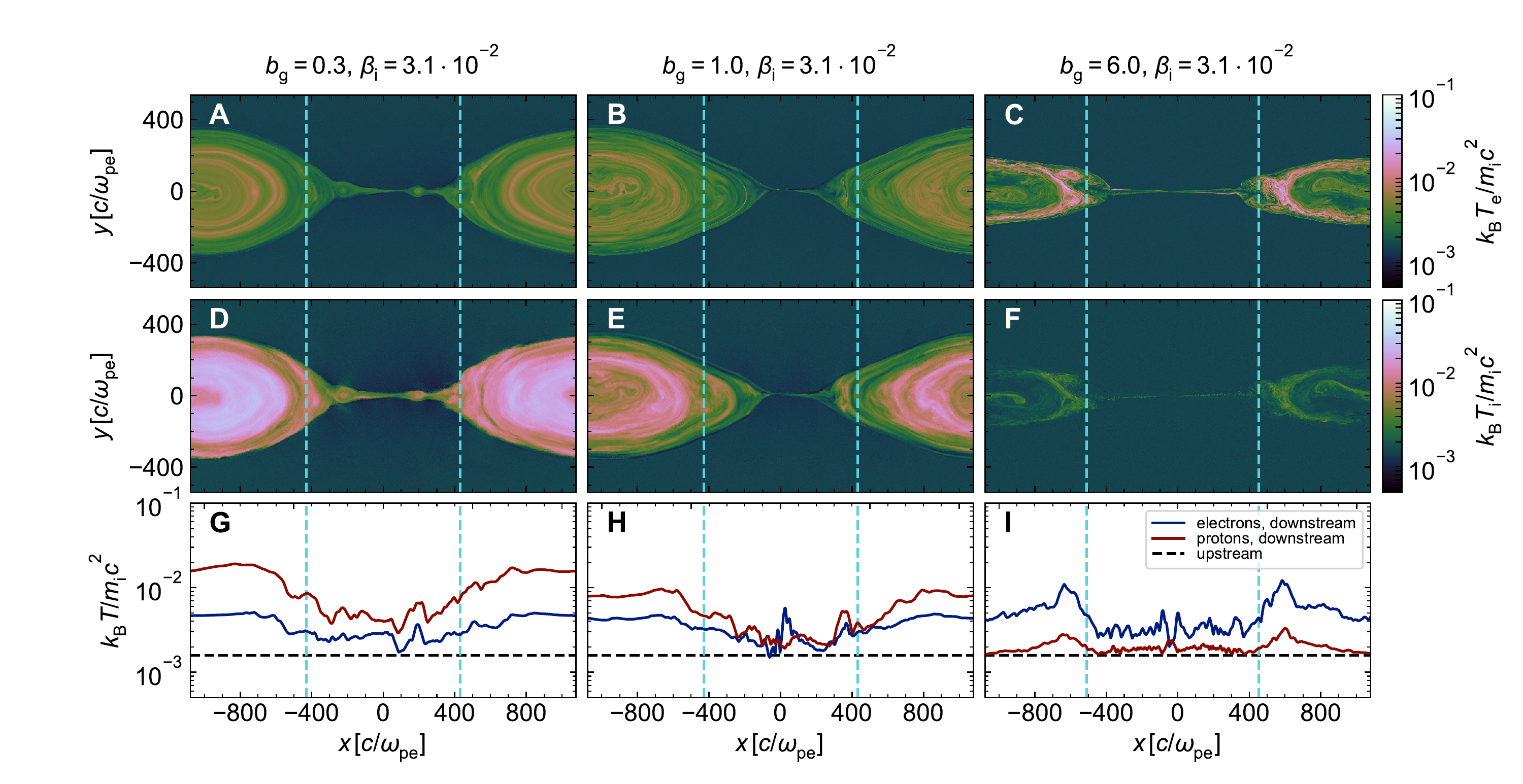} \\
		\caption{
		Comparison of electron and proton heating for guide fields $\bg=0.3$ (first column), $\bg=1$ (second column), and $\bg=6$ (third column); $\betai \approx 0.03$ for these simulations, which correspond to \texttt{b3e-2.bg3e-1}, \texttt{b3e-2.bg1}, and \texttt{b3e-2.bg6} in \tab{params}.  The first, second, and third rows show 2D plots of electron temperature, 2D plots of proton temperature, and 1D profiles (averaged along $y$, for cells in the downstream) of electron (blue) and proton (red) temperature.  In the bottom row, the dashed black line shows the initial temperature in the upstream.  Vertical cyan dashed lines indicate the $x$ boundaries of the island region; no cells between the cyan lines are counted as part of the island region.  Note that the electron and proton temperatures are both normalized to $m_{\rm i} c^2$.  The snapshots are shown at time $t/\ta = 2.7$ (equivalently, $\tomp \approx 2 \times 10^4$).  1D profiles are slightly smoothed for clarity.  An animated version of this figure is available from the online journal.  \label{fig:gf_compare2} \\} 
\end{figure*}
\begin{figure*}
		\centering
		\includegraphics[width=\textwidth,clip,trim=0cm 0cm 0cm 0.0cm]{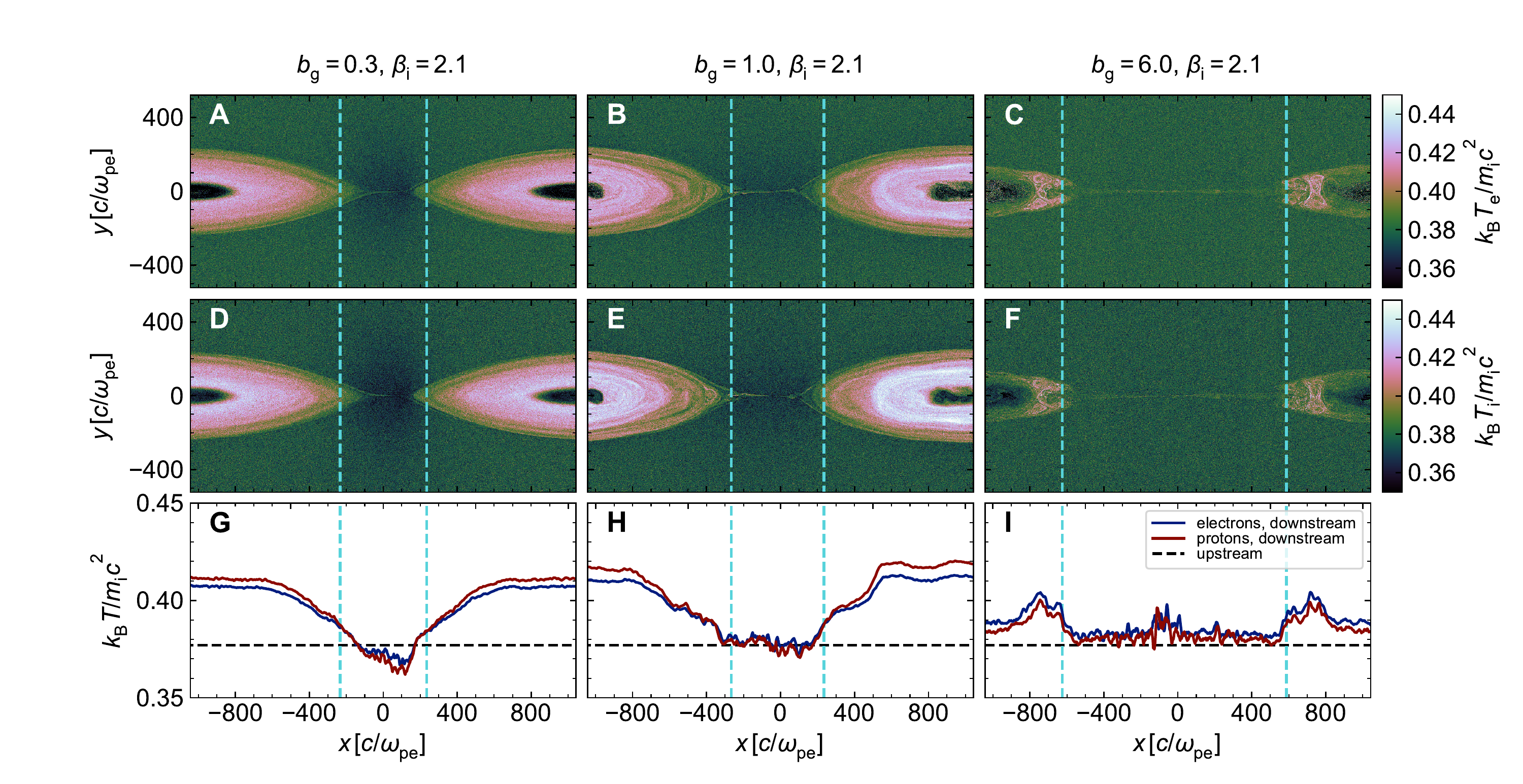} \\
		\caption{
		The layout is similar to \fig{gf_compare2}.  Comparison of electron and proton heating for guide fields $\bg=0.3$ (first column), $\bg=1$ (second column), and $\bg=6$ (third column), but for $\betai \approx 2$;  these are simulations \texttt{b2.bg3e-1}, \texttt{b2.bg1}, and \texttt{b2.bg6} in \tab{params}.  The first, second, and third rows show 2D plots of electron temperature, 2D plots of proton temperature, and 1D profiles of electron (blue) and proton (red) temperature; here, temperatures are normalized to $m_{\rm i} c^2$.  The meaning of dashed black and cyan lines is as in \fig{gf_compare2}.  The snapshots are at time $t/\ta = 3.1$ ($\tomp \approx 2.2 \times 10^4$). 1D profiles are slightly smoothed for clarity.  An animated version of this figure is available from the online journal.  \label{fig:gf_compare3} \\} 
\end{figure*}
Electron and proton heating via reconnection shows substantial differences in the limits of strong and weak guide field.  
\fig{gf_compare2} shows 2D snapshots at $t/\ta=2.7$ of electron (panels A--C) and proton (panels D--F) temperature,\footnote{Here, we phrase our results in terms of temperature, rather than internal energy; however, similar conclusions hold regardless of which quantity is considered (in this section as well as in the rest of the paper).} and corresponding 1D profiles (panels G--I), for three simulations with a relatively low $\betai=0.03$ and guide field strengths $\bg=0.3, 1,$ and $6$, increasing from left to right.  The simulations here correspond to runs \texttt{b3e-2.bg3e-1}, \texttt{b3e-2.bg1}, and \texttt{b3e-2.bg6} in \tab{params}.

The first and second rows show the spatial dependence of electron (panels A--C) and proton (panels D--F) heating.  At low $\bg$ (run \texttt{b3e-2.bg3e-1}), the electron and proton temperatures are relatively uniform in the exhaust and island regions.  For intermediate guide field strengths (run \texttt{b3e-2.bg1}), the electron and proton heating is less uniform in the island, and shows marked a asymmetry in the exhaust region (see panels B and E, in between the cyan lines).  For the strong guide field case (run \texttt{b3e-2.bg6}), electrons reach a maximum temperature of roughly $k_{\rm B} T_{\rm e}/m_{\rm i} c^2 \approx 0.02$ along the upper-left and lower-right edges of the outflow; on the other hand, proton heating along the exhaust is essentially isolated to the upper-right and lower-left edges.  Throughout the entire downstream (for run \texttt{b3e-2.bg6}), the proton temperature rarely exceeds $k_{\rm B} T_{\rm i}/m_{\rm i} c^2 \approx 5 \times 10^{-3}$.  

The 2D plots in panels A--F also illustrate that the primary island becomes more oblong with increasing guide field. For $\bg=0.3$, the aspect ratio of the island (length along the layer to width orthogonal to it)  is about $7:4$, whereas at $\bg=6$, it is twice as large, $7:2$.  In the cases with strong guide field the primary islands do not circularize up to the run times of our simulations.

The bottom row of \fig{gf_compare2} shows the 1D profiles of electron (blue) and proton (red) temperatures, both in units of $m_{\rm i} c^2$, averaged along $y$ for cells within the downstream region (including the yellow and tan regions in panel C of \fig{rec_isl}, and excluding the grey area in the island core that contains particles left over from initialization).  
The edges of the primary island are shown by vertical cyan lines.  Horizontal black dashed lines indicate the initial temperature of particles in the far upstream.  In the weak guide field case $\bg=0.3$ (panel G),  protons are heated substantially more than  electrons, similar to the case of antiparallel reconnection (see \citet{Melzani2014}, \citet{Werner2016}, \citetalias{rsn17}).  As the strength of the guide field increases (panels H and I),  proton heating in both the exhaust region and the primary island  is strongly suppressed.  The electron temperature, on the other hand, is largely unaffected; for $\bg=0.3, 1,$ and $6$, the electron temperature in the island is always around $k_{\rm B} T_{\rm e}/m_{\rm i} c^2 \approx 5 \times 10^{-3}$.

\fig{gf_compare3} is similar to \fig{gf_compare2}, but corresponds to a set of simulations with $\betai=2$ (runs \texttt{b2.bg3e-1}, \texttt{b2.bg1}, and \texttt{b2.bg6} in \tab{params}). As we discuss below, this value of $\betai=2$ is close to $\beta_{\rm i,max}=2.5$ (see \eq{betaimax}), impliying that electrons and protons both start with relativistic temperatures. 
 In stark contrast to the low $\betai$ case, at $\betai=2$ the electron and proton temperatures in the island region are roughly equal, regardless of the guide field strength ($\bg = 0.3$--$6$).  Still, the 2D temperature structure within the island differs between low and high guide field cases.  At high $\betai$ and low or intermediate guide field (runs \texttt{b2.bg3e-1} and run \texttt{b2.bg1}), the electron and proton temperatures in the island are typically uniform (similar to the low $\betai$, low $\bg$ case in panels A and D of \fig{gf_compare2}). However, at high $\betai$ and high guide field (run \texttt{b2.bg6}), the electron and proton temperatures are less uniform (relative to runs \texttt{b2.bg3e-1} and \texttt{b2.bg1}; see panels C and F of \fig{gf_compare2}), with electron and proton temperatures greatest near the interfaces between the primary island and the outflows (i.e., $x=\pm 700 \comp$).

\subsection{Reconnection Rate}\label{sec:recrate}
\begin{figure}
		\centering
		\includegraphics[width=0.45\textwidth,clip,trim=0cm 0cm 0cm 0.0cm]{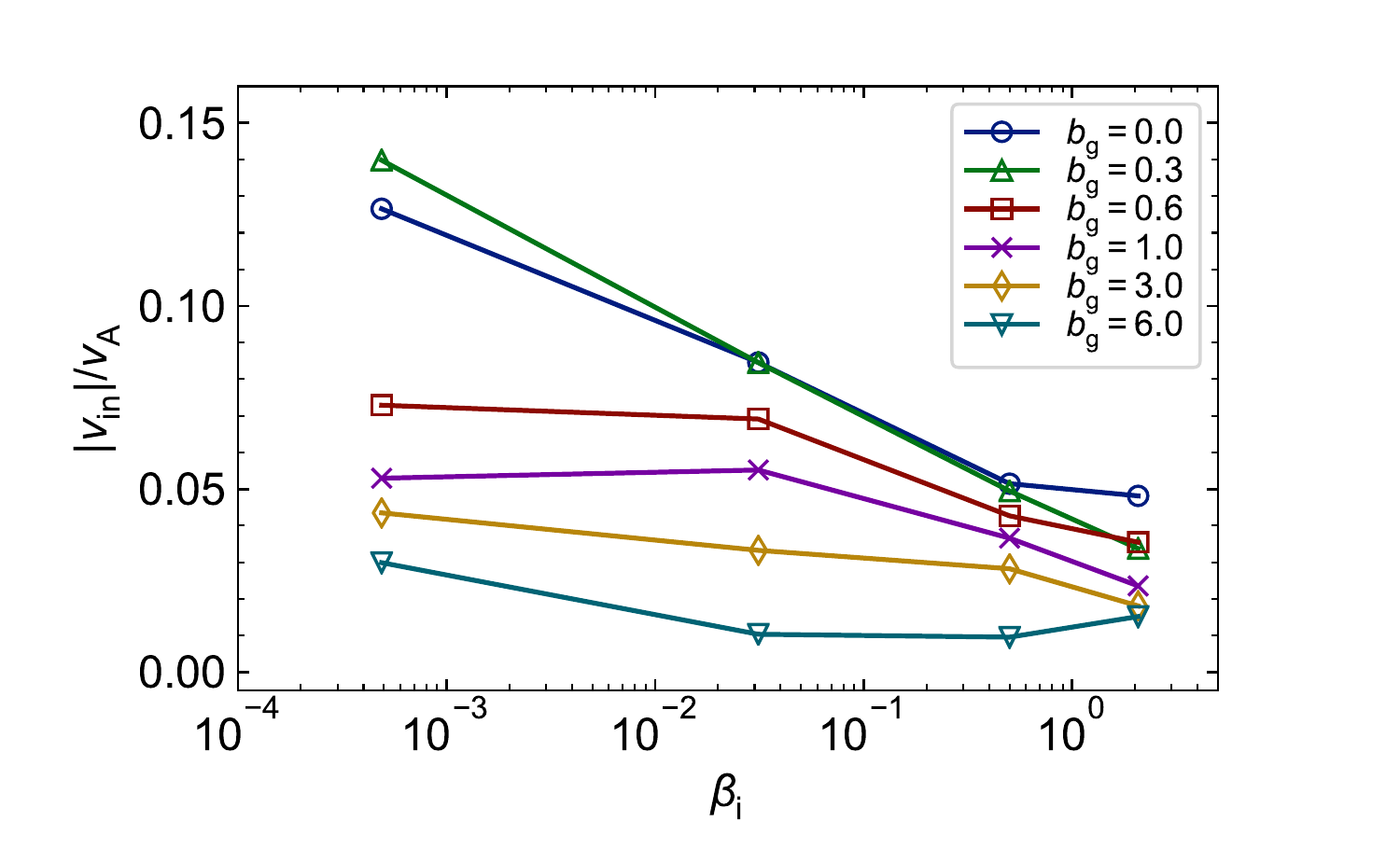} \\
		\caption{Reconnection rate, i.e., the upstream inflow velocity in units of the Alfv\'{e}n velocity, for the main simulations in \tab{params}.  Blue, green, red, purple, yellow, and teal points indicate simulations with guide fields $0, 0.3, 0.6, 1, 3$, and $6$ (solid lines are included to guide the eye).  The inflow velocity is averaged in time from $t/\ta \approx 0.7$ to $1$ ($\tomp$ from 5000 to 7100), and spatially over the upstream region (similar to what is delimited in \fig{rec_isl} by the dashed white lines, but see text for details). \label{fig:vinvout_bth} \\} 
\end{figure}

\fig{vinvout_bth} shows the $\betai$ and $\bg$ dependence of the reconnection rate, $|v_{\rm in}| /v_{\rm A}$.  The inflow speed $|v_{\rm in}|$ is computed as a spatial average over a specific region of the upstream,\footnote{Specifically, in the region where $100 \comp < |y| < 120 \comp$ and $|x|< 360 \comp$.} and temporal average from $t/\ta \approx 0.7$ to $1$ (when reconnection is roughly in steady state).  Each point corresponds to the measurement from a different simulation, and those with the same guide field strength $\bg$ are connected by a solid line.  For these simulations, $\mime=1836$, $\teti=1$, and $\sigmaw=0.1$.

In most cases, reconnection proceeds at or below the value often reported in the literature, i.e. $|v_{\rm in}| /v_{\rm A} \lesssim 0.1$ \citep{Cassak17}; however, for low $\betai$ and weak guide field ($\bg \lesssim 0.3$), the reconnection rate exceeds this fiducial value, with $|v_{\rm in}| /v_{\rm A}$ in the range 0.1--0.15.  For $\bg \lesssim 0.3$, the reconnection rate shows a relatively weak scaling with $\betai$, decreasing from $|v_{\rm in}|/v_{\rm A} \approx 0.1$--$0.15$ to $|v_{\rm in}|/v_{\rm A}\approx 0.05$, only a factor of 2--3, as $\betai$ increases from $5 \times 10^{-4}$ to $2$ (\citealp{Numata15}, \citetalias{rsn17}, \citealp{Ball2018}).  For guide fields $\bg \gtrsim 1$, the $\betai$ dependence of the reconnection rate is even weaker, and $|v_{\rm in}|/v_{\rm A}$ typically varies from 0.01 to 0.07.  We find that the presence of a guide field tends to suppress the reconnection rate, which is a dependence similar to that found by \citet{Melzani2014} for electron-ion relativistic reconnection, \citet{Ricci2003}, \citet{Huba2005}, \citet{TenBarge2013}, and \citet{Liu2014} for electron-ion nonrelativistic reconnection, and \citet{Hesse2007} and \citet{Werner2017} for electron-positron plasma.  The decrease in the reconnection rate with $\bg$ is more pronounced at lower values of $\betai$.

\begin{figure}
		\centering
		\includegraphics[width=0.45\textwidth,clip,trim=0cm 2cm 0cm 1.5cm]{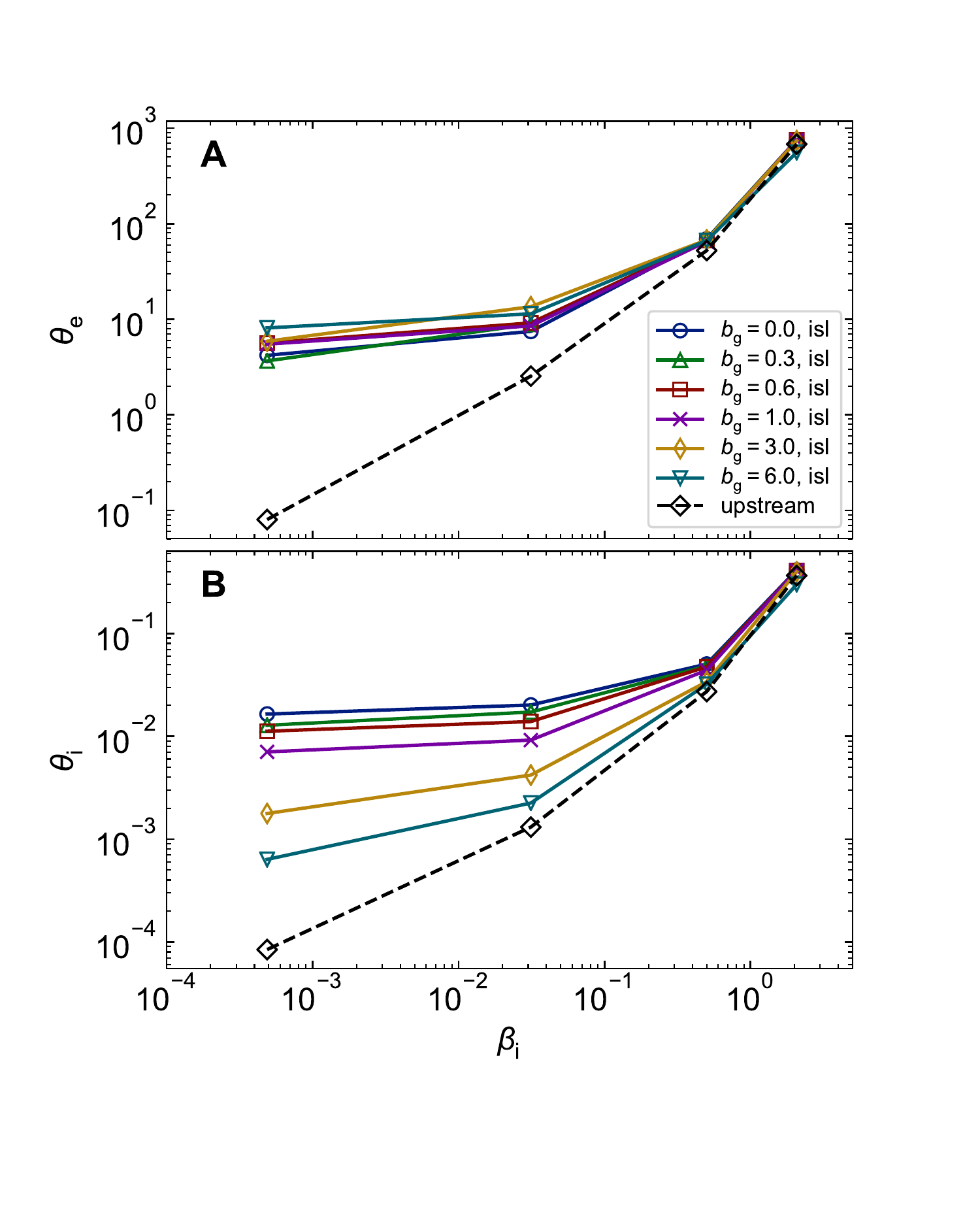} \\
		\caption{Dimensionless (A) electron and (B) proton temperatures, in units of $m_{\rm e} c^2$ and $m_{\rm i} c^2$, respectively, measured in the island region (yellow region in \fig{rec_isl}).  The color scheme is the same as in \fig{vinvout_bth}, with different colors indicating simulations with different guide field strength.  The black diamond points show the temperature in the upstream region.  The measured temperatures are averaged over $\sim 1\ta$ (equivalently, $\sim7100\omega_{\rm pe}^{-1}$). \bigskip \label{fig:dindout_bth} \\} 
\end{figure}

\subsection{Electron and Proton Heating: $\bg$ and $\betai$ Dependence}\label{sec:gfheating}

\fig{dindout_bth} shows the $\bg$ and $\betai$ dependence of electron (panel A) and proton (panel B) dimensionless temperature.  Each solid line shows the volume-averaged temperature in the island for a set of simulations with the same value of $\bg$, and the black diamonds, connected with a dashed line, show the upstream temperature for each value of $\betai$ (for simulations with fixed $\betai$, the upstream temperature is the same, independent of $\bg$).  As discussed in \sect{setup}, the numerical resolution is sufficient to keep numerical heating under control; temperatures measured in the inflow region are about the same as the ones at initialization, throughout the duration of our simulations.\footnote{Due to numerical heating, the measured upstream temperature $\theta_{\rm e} \approx 0.08$ for the $\betai=5 \times 10^{-4}$ runs differs from the expected initialized electron temperature, $\theta_{\rm e0}=0.045$ (see \tab{params}).  However, the difference is much smaller (less than ${\sim}15\%$) in runs with higher $\betai$.  Although the upstream numerical heating for $\betai=5 \times 10^{-4}$ is large compared to the temperature at initialization, the resulting upstream temperature is still much smaller than the downstream temperature, so it has no effect on the heating fractions presented below.  The basic reason is that for low $\betai$, the available magnetic energy (a fraction of which will be transferred to the particles) is much larger than the initial particle thermal energy.}   The simulations presented here are the same as those in \sect{recrate}, so they employ $\mime=1836$, $\teti=1$, and $\sigmaw=0.1$. 

The upstream electron dimensionless temperatures range from nonrelativistic, $\theta_{\rm e} \approx 0.08$, up to ultra-relativistic, $\theta_{\rm e} \approx 700$; the temperatures in the downstream region range from moderately to ultra-relativistic, $\theta_{\rm e} \approx 6 $--$700$.\footnote{This is a consequence of our choice of $\sigmaw=0.1$; for $\sigmaw \ll1$, the bulk of electrons will not attain ultra-relativistic energies.}
For all values of $\betai$, the electron temperature in the island appears to be nearly independent of the guide field strength (\fig{dindout_bth}, panel A).  The guide field simulations show the same scaling with $\betai$ as  the antiparallel case (blue circles), with the electron temperature  increasing from about $\theta_{\rm e} \approx 6$ at $\betai \sim 5 \times 10^{-4} $ up to $\theta_{\rm e} \approx 700$ at $\betai \sim 2$.  Additionally, the electron temperature shows only a relatively weak dependence on $\betai$, for $\betai \lesssim 0.5$.  From $\betai \sim 5 \times 10^{-4}$ up to $0.5$, the electron temperature in the island changes by no more than a factor of 10 (the dependence is even weaker for $\betai\lesssim3\times 10^{-2}$).  At high $\betai$, the electron temperature in the island appears to be nearly the same as that in the upstream. However, the increase in temperature from upstream to downstream corresponds to a substantial fraction (typically ${\sim}30\%$) of the inflowing magnetic energy per electron (see \fig{mu_bth} below, panel A).  These two statements are not in contradiction, since for high $\betai$ the available magnetic energy is only a small fraction of the initial thermal energy.  

In \fig{dindout_bth}, panel B, we show the proton dimensionless temperature in the island, for the same simulations shown in panel A.  At low $\betai$, protons show a clear decrease in island temperature with increasing guide field strength; for antiparallel reconnection ($\bg=0$) and $\betai \sim 5 \times 10^{-4}$, the proton dimensionless temperature in the island is $\theta_{\rm i} \approx 0.02,$ but decreases to $\theta_{\rm i} \approx 6 \times 10^{-4}$ for strong guide field, $\bg=6$.  As $\betai$ increases, the proton heating in the island shows a weaker dependence on guide field strength.  Similar to electron heating in the island at high $\betai$, $\theta_{\rm i}$ is nearly independent of $\bg$ at high $\betai$.  The proton dimensionless temperatures, in both the upstream and the island region, are generally nonrelativistic, $\theta_{\rm i} \lesssim 1$.  In summary, when comparing panels A and B, a striking difference is that the electron dimensionless temperature in the island is independent of the guide field strength, whereas the proton dimensionless temperature appreciably decreases with increasing $\bg$.  

In \fig{mu_bth} we present the scaling of electron and proton heating with guide field strength $\bg$ and proton-$\betai$. The first and second rows show the electron and proton heating fractions, respectively (see Eqs. \ref{eq:mue}--\ref{eq:muidecomp}); the total heating (first column) is decomposed into adiabatic-compressive and irreversible components, shown in the second and third columns, respectively.  In each panel, the corresponding heating fraction is plotted as a function of $\betai$ for guide field strengths in the range $0$--$6$.  

The first row in \fig{mu_bth} shows the scaling of the electron total, adiabatic, and irreversible heating fractions ($\muetot, \muead$, and $\mueirr$) with respect to $\bg$ and $\betai$.  At low $\betai$, the electron total heating fraction within the island does not show a strong scaling with the strength of the guide field (consistent with \fig{dindout_bth}).  For $\betai \lesssim 0.03$, $\muetot \sim 0.1$.  At high $\betai$, the total heating is suppressed by strong guide fields, $\bg \gtrsim 3$.  Some insight into this trend is provided by decomposing the total heating fraction $\muetot$ into adiabatic and irreversible parts, $\muead$ (panel B) and $\mueirr$ (panel C).  For low $\betai$, compressive heating is negligible; however, at higher values of $\betai$, compressive heating is more significant, but tends to decrease with stronger guide fields, which is in qualitative agreement with \citet{Li2018}.  This result is physically intuitive, as the plasma becomes less compressible when the magnetic pressure of the guide field is larger (and in fact, we notice that the primary island is less dense for stronger guide fields).  

To summarize, we find that the electron compressive heating fraction in panel (B) steadily increases with $\betai$ and strongly decreases with $\bg$.  Both trends for $\muead$ can be easily understood from \eq{compapprox}, given that stronger guide fields give smaller density compressions. In contrast, the electron irreversible heating fraction (panel C) is largely independent of both $\bg$ and $\betai$, and it is around $\mueirr\sim0.1$.  The combination of irreversible and compressive heating explains why the total heating at low $\betai$ is independent of both $\betai$ and $\bg$, whereas at high $\betai$ it is lower for larger $\bg$ (due to the corresponding trend in compressive heating). 

The second row in \fig{mu_bth} shows the proton heating fractions $\muitot, \muiad$, and $\muiirr$ (panels D, E, and F).  The proton total heating in the island differs sharply from the electron total heating (panel A).  The proton total heating shows a strong dependence on the strength of the guide field; for antiparallel reconnection, $\muitot \approx 0.3$ regardless of $\betai$, but the total heating is significantly suppressed as $\bg$ increases.  For $\bg=6$ and $\betai \lesssim 0.5$, $\muitot$ is negligible.  The proton compressive heating (panel E) shows a trend similar to that of the electron compressive heating (panel B); for both electrons and protons, the compressive heating is controlled by density in the upstream, density in the island region, and upstream temperature (here, we focus on the case $\teti=1$);  since these quantities are similar for electrons and protons, the compressive heating for both species shows the same trend.  The proton irreversible heating (panel F) is similar to the proton total heating (panel A) for $\betai \lesssim 0.03$, because  compressive heating is negligible in this regime.  For $\betai \gtrsim 0.03$, the proton irreversible heating is less sensitive to the guide field strength, and by $\betai=2$, $\muiirr \approx 0.08$ regardless of $\bg$, similarly to the electron irreversible heating.

The electron and proton irreversible heating fractions $\mueirr$ and $\muiirr$ can be used to compute the ratio of electron irreversible heating to total irreversible particle heating liberated during reconnection \citepalias{rsn17},
\bea \label{eq:queirrev}
	q_{u \rm e, irr} \equiv \frac{\mueirr}{\mueirr + \muiirr}.
\eea
In \fig{que_irr_bth}, we present the $\betai$ and $\bg$ dependence of $\queirr$, the \textit{electron irreversible heating efficiency}.  For all $\betai \lesssim 2$, $\queirr$ increases with the guide field strength.  For antiparallel reconnection, electrons ultimately receive ${\sim}18\%$ of the irreversible heat transferred to particles.  As the guide field increases, so does the fraction of irreversible heating transferred to electrons; for $\bg=1, \queirr \approx 45\%$, and by $\bg=6$, electrons receive the vast majority of magnetic energy that is converted to irreversible particle heating, with $\queirr \approx 93\%$.  At $\betai=2\sim\beta_{\rm i, max}$, $\queirr\approx 50\%$, independently of $\bg$; $\beta_{\rm i, max}$ is the maximum possible value of $\betai$, given $\sigmaw$ and $\teti$, and is defined as
\bea \label{eq:betaimax}
\beta_{\rm i, max} = \frac{0.5}{\sigmaw + \sigmaw \teti}.
\eea 
This equation is derived by expressing $\betai$ as a function of $\teti, \sigmaw$, and $\theta_{\rm i}$, then taking the limit $\theta_{\rm i} \rightarrow \infty$.  For the simulations presented here, with $\mime=1836$, $\teti=1$, and $\sigmaw=0.1$, we find $\beta_{\rm i, max}=2.5$.  Note that for $\betai \sim \beta_{\rm i, max}$, electrons and protons start relativistically hot in the upstream, and the scale separation $(\comp)/(\compi)$ is of order unity \citepalias{rsn17}; in this case, electrons and protons behave nearly the same, which explains why for $\betai \sim \beta_{\rm i, max}$ we obtain energy equipartition, i.e., we find that $\queirr\approx 50\%$, independently of $\bg$.

\onecolumngrid
\begin{center}
\begin{figure}[h]
		\centering
		\includegraphics[width=\textwidth,clip,trim=0cm 0.5cm 0cm 0.0cm]{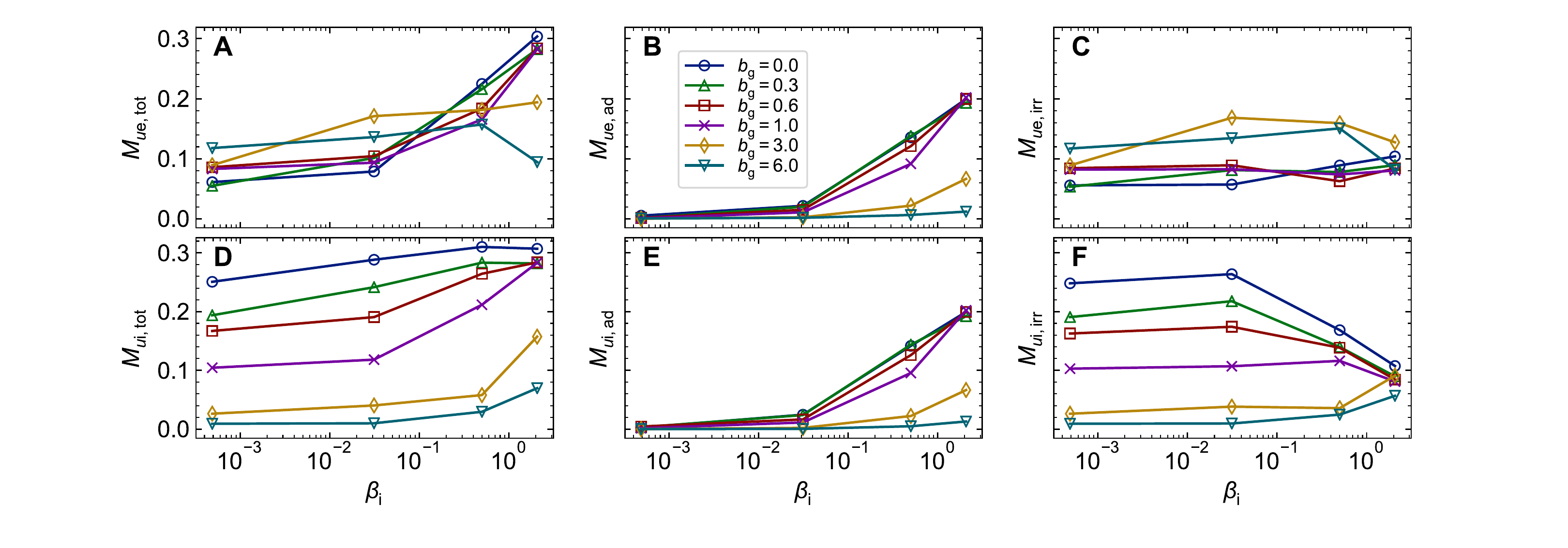} \\
		\caption{
		Guide field $\bg$ and proton-$\betai$ dependence of electron (A) total, (B) adiabatic, (C) irreversible heating; proton (D) total, (E) adiabatic, and (F) irreversible heating.  The heating fractions are defined in \sect{methods} (see Eqs. \ref{eq:mue}--\ref{eq:muidecomp}).  For these simulations, $\mime=1836$, $\sigmaw=0.1$, and $\teti=1$.\label{fig:mu_bth} \\} 
\end{figure}
\end{center}
\twocolumngrid

\begin{figure}[H]
		\centering
		\includegraphics[width=0.45\textwidth,clip,trim=0cm 0cm 0cm 0.0cm]{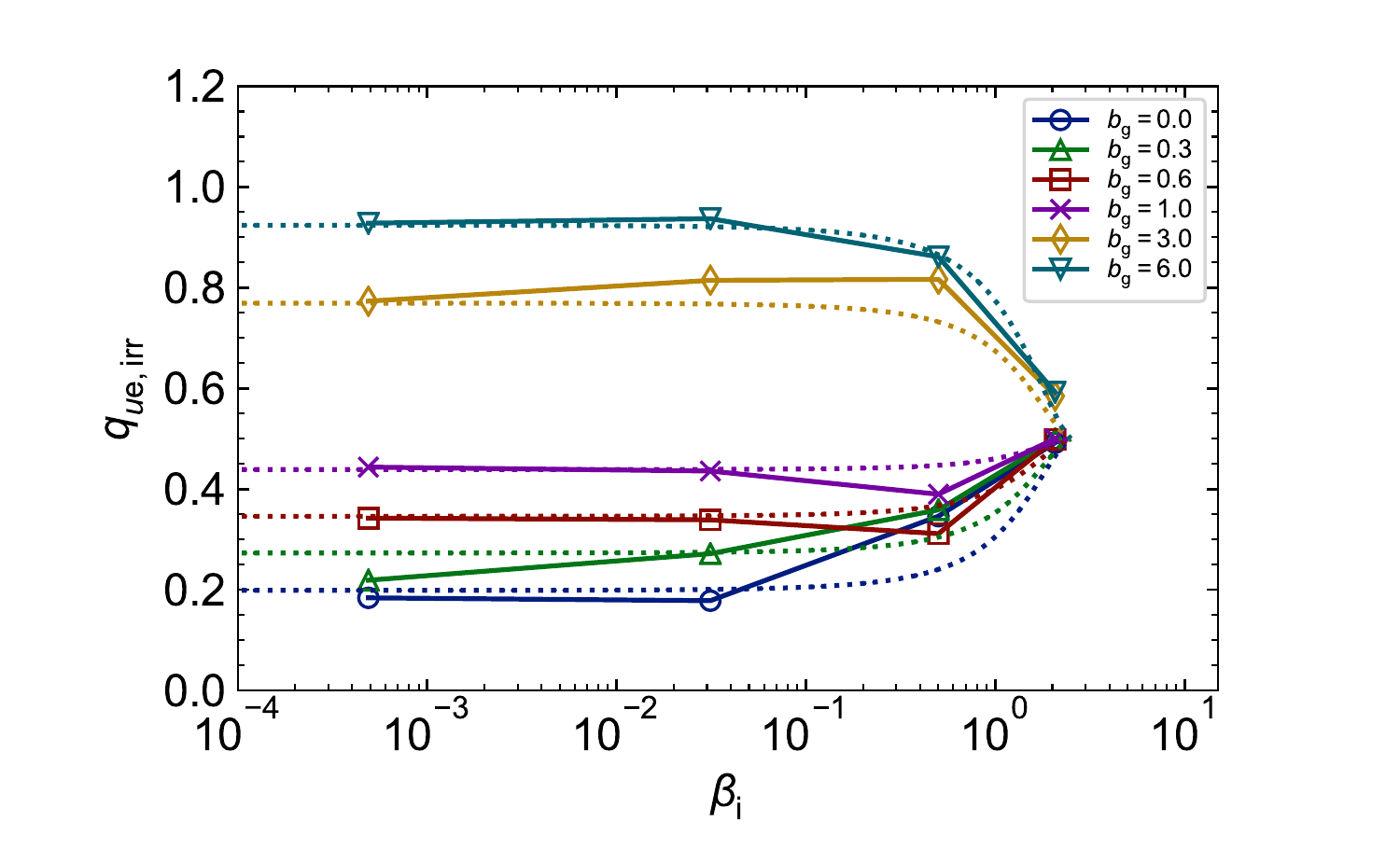} \\
		\caption{Guide field and proton-$\betai$ dependence of electron irreversible heating efficiency, $\queirr$ (see \eq{queirrev}).  The values plotted here are computed from $\mueirr$ and $\muiirr$ shown in panels C and F of \fig{mu_bth}.  Dotted lines show the fitting function in \eq{fit} for $\bg$ in the range 0--6.\label{fig:que_irr_bth} \\} 
\end{figure}

\subsection{Electron Irreversible Heating Efficiency: $T_{\rm e0}/T_{\rm i0}$ and $\sigma_{w}$ Dependence}\label{sec:others}
\begin{figure}[H]
		\centering
		\includegraphics[width=0.45\textwidth,clip,trim=0cm 3cm 0cm 2cm]{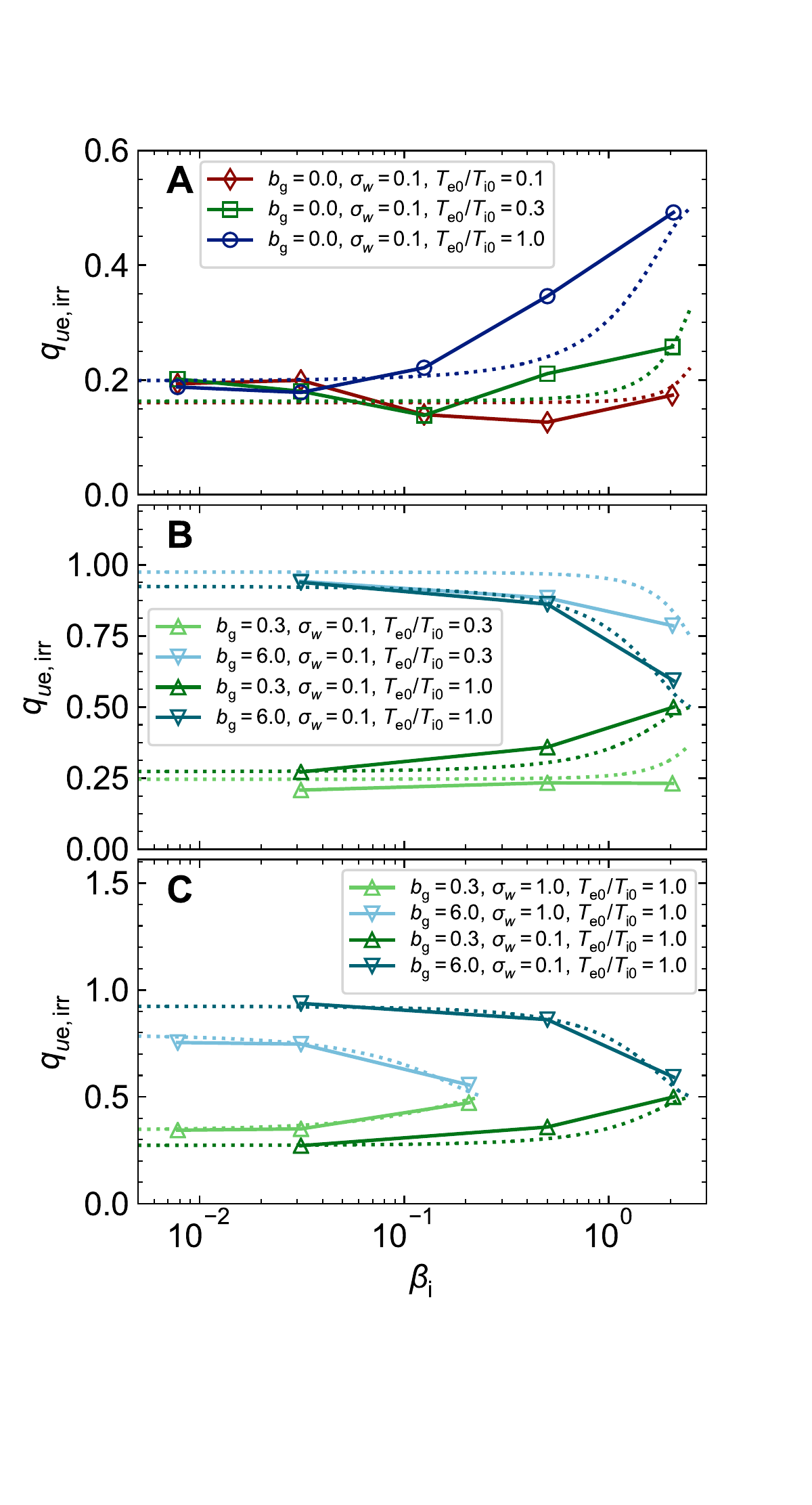} \\
		\caption{Similar to \fig{que_irr_bth}, but for the simulations listed in \tab{params2} rather than \tab{params} (fiducial cases with $\sigmaw=0.1, \teti=1$ are also shown for reference); dependence of electron irreversible heating efficiency, $\queirr$, for (A) $\teti=0.1$ up to $1$ and antiparallel reconnection, (B) unequal initial electron and proton temperatures in the upstream, $\teti=0.3$, for two guide field cases, $\bg=0.3$ and $6$, and (C) $\sigmaw=1$, again for $\bg=0.3$ and $6$.  As before, dotted lines show the fitting function in \eq{fit}. \bigskip \label{fig:que_irr_others} \\} 
\end{figure}

For simplicity, we focused in \sect{gfheating} on electron heating for cases with representative magnetization $\sigmaw=0.1$ and temperature ratio $\teti=1$.  A full exploration of the dependence of electron and proton heating on $\betai$, $\bg$, $\sigmaw$, and $\teti$ is beyond the scope of this work.  Nevertheless, for a limited range of $\bg$ and $\betai$, we present in \fig{que_irr_others} the electron irreversible heating efficiencies when we vary the electron-to-proton temperature ratio $\teti$ in the range 0.1--1 (panels A and B), as well as for several simulations with $\sigma_{w}=1$ (panel C).  The physical parameters of these runs are given in \tab{params2}.

The effect of varying the initial electron-to-proton temperature ratio for antiparallel reconnection ($b_{\rm g}=0$) is demonstrated in panel A of \fig{que_irr_others}.  At low $\betai$, the electron irreversible heating efficiency shows nearly no dependence on $\betai$ or temperature ratio.  At high $\betai$, the dependence on temperature ratio can be understood via the dependence of $\beta_{\rm i, max}$ on $T_{\rm e0}/T_{\rm i0}$.  According to \eq{betaimax}, decreasing the temperature ratio for fixed $\sigmaw$ leads to an increase in $\beta_{\rm i, max}$, and so (as discussed in \sect{recrate}) in the value of $\betai\sim \beta_{\rm i, max}$ where equipartition between electrons and protons is realized.

The effect of varying the temperature ratio for $\bg=0.3$ and $\bg=6$ is shown in panel B.  As for antiparallel reconnection, there is no significant dependence on $T_{\rm e0}/T_{\rm i0}$ at low $\betai$, for each of the two $\bg$ values. While $\beta_{\rm i, max}=2.5$ for $\teti=1$, for $\teti=0.3$ we expect $ \beta_{\rm i, max}\approx3.85$, so equipartition between electrons and protons, which should hold regardless of $\bg$ at $\betai\sim \beta_{\rm i, max}$, is expected at higher $\betai$ than probed in panel (B). 

The effect of varying the magnetization and guide field strength is shown in panel C of \fig{que_irr_others}.  At low $\betai$, 
the electron irreversible heating efficiency has a weaker dependence on guide field for $\sigmaw=1$ than for $\sigmaw=0.1$.  For $\betai \sim \beta_{\rm i, max}\propto \sigma_w^{-1}$ (see \eq{betaimax}), irreversible heating of electrons and protons is in equipartition, and this conclusion holds regardless of $\sigmaw$ or $\bg$.

\begin{table*}
\begin{tabularx}{\linewidth}{c X}
  \toprule \midrule
 \makecell{\textbf{Run ID:} \\ $\betai$ \\ $\bg$ \\ $\sigmaw$ \\ $\teti$} 
 			& \hfill\makecell{\texttt{b8e-3.bg0.t1e-1} \\ $7.8 \times 10^{-3}$ \\ 0 \\ 0.1 \\ 0.1}
                             \hfill\makecell{\texttt{b3e-2.bg0.t1e-1} \\ 0.031                       \\ 0  \\ 0.1 \\ 0.1}
                             \hfill\makecell{\texttt{b1e-1.bg0.t1e-1} \\ 0.13                         \\ 0  \\ 0.1 \\ 0.1}
                             \hfill\makecell{\texttt{b5e-1.bg0.t1e-1} \\ 0.5                           \\ 0 \\ 0.1 \\ 0.1}
                             \hfill\makecell{\texttt{b2.bg0.t1e-1}       \\ 2                             \\ 0 \\ 0.1 \\ 0.1} \hfill\null \\     
               \midrule
 \makecell{\textbf{Run ID:} \\ $\betai$ \\ $\bg$ \\ $\sigmaw$ \\ $\teti$} 
 			& \hfill\makecell{\texttt{b8e-3.bg0.t3e-1} \\ $7.8 \times 10^{-3}$ \\ 0 \\ 0.1 \\ 0.3}
                             \hfill\makecell{\texttt{b3e-2.bg0.t3e-1} \\ 0.031                       \\ 0  \\ 0.1 \\ 0.3}
                             \hfill\makecell{\texttt{b1e-1.bg0.t3e-1} \\ 0.13                         \\ 0  \\ 0.1 \\ 0.3}
                             \hfill\makecell{\texttt{b5e-1.bg0.t3e-1} \\ 0.5                           \\ 0 \\ 0.1 \\ 0.3}
                             \hfill\makecell{\texttt{b2.bg0.t3e-1}       \\ 2                             \\ 0 \\ 0.1 \\ 0.3} \hfill\null \\     
               \midrule
 \makecell{\textbf{Run ID:} \\ $\betai$ \\ $\bg$ \\ $\sigmaw$ \\ $\teti$} 
 			& \hfill\makecell{\texttt{b3e-2.bg3e-1.t3e-1} \\ 0.031 \\ 0.3 \\ 0.1 \\ 0.3}
                             \hfill\makecell{\texttt{b5e-1.bg3e-1.t3e-1} \\ 0.5 \\ 0.3 \\ 0.1 \\ 0.3}
                             \hfill\makecell{\texttt{b2.bg3e-1.t3e-1} \\ 2 \\ 0.3 \\ 0.1 \\ 0.3}
                             \hfill\makecell{\texttt{b3e-2.bg6.t3e-1} \\ 0.031 \\ 6 \\ 0.1 \\ 0.3}
                             \hfill\makecell{\texttt{b5e-1.bg6.t3e-1} \\ 0.5 \\ 6 \\ 0.1 \\ 0.3}
                             \hfill\makecell{\texttt{b2.bg6.t3e-1} \\ 2 \\ 6 \\ 0.1 \\ 0.3} \hfill\null \\
     \midrule
 \makecell{\textbf{Run ID:} \\ $\betai$ \\ $\bg$ \\ $\sigmaw$ \\ $\teti$} 
 			& \hfill\makecell{\texttt{b8e-3.bg3e-1.s1} \\ $7.8 \times 10^{-3}$ \\ 0.3 \\ 1 \\ 1}
                             \hfill\makecell{\texttt{b3e-2.bg3e-1.s1} \\ 0.031 \\ 0.3 \\ 1 \\ 1}
                             \hfill\makecell{\texttt{b2e-1.bg3e-1.s1} \\ 0.2 \\ 0.3 \\ 1 \\ 1}
                             \hfill\makecell{\texttt{b8e-3.bg6.s1} \\ $7.8 \times 10^{-3}$ \\ 6 \\ 1 \\ 1}
                             \hfill\makecell{\texttt{b3e-2.bg6.s1} \\ 0.031 \\ 6 \\ 1 \\ 1}
                             \hfill\makecell{\texttt{b2e-1.bg6.s1} \\ 0.2 \\ 6 \\ 1 \\ 1} \hfill\null \\
  \bottomrule
\end{tabularx}
\caption{\raggedright Physical parameters for simulations with unequal temperature ratios, as well as $\sigma_{w}=1$, described in \sect{others}.}
\label{tab:params2}
\end{table*}

\subsection{Fitting function}\label{sec:fitfunc}
For use as a sub-grid model of electron heating in magnetohydrodynamic simulations (as in \citet{Ressler17, Chael18, Ryan18}), we provide the following fitting formula, motivated by the simulation results presented in Secs. \ref{sec:gfheating} and \ref{sec:others}:  
\bea
\begin{split} \label{eq:fit}
q_{u \rm e, irr, fit}(\betai, \bg, \teti, &\sigmaw) = \frac{1}{2}\left(\tanh\left(0.33 \,  \bg\right) - 0.4 \right) \\
\times 1.7\tanh &\left(\frac{\left(1 - \betai/\beta_{\rm i, max}\right)^{1.5}}{(0.42 + \teti) \sigmaw^{0.3}}\right) + \frac{1}{2},
\end{split}
\eea
where $\beta_{\rm i, max}$ is in \eq{betaimax} in terms of $\sigma_w$ and $T_{\rm e0}/T_{\rm i0}$.

The fitting function in \eq{fit} has the following limits: for low $\betai$, $q_{u \rm e, irr, fit}$ asymptotes to a ($\sigmaw$- and $b_{\rm g}$-dependent) value that does not depend on $\betai$. The asymptotic low-$\betai$ limit tends to the equipartition value $q_{u \rm e, irr, fit}\sim 0.5$ for $\sigma_w\gg1$ (i.e., in the limit of ultra-relativistic reconnection), regardless of $\bg$. Still at $\betai\ll1$, electrons receive most of the irreversible heat if 
 $\bg\gtrsim1.3$. For $\bg\gg1$, $\sigma_w\ll1$ and $\betai\ll1$, we get $q_{u \rm e, irr, fit}\approx1.0$, i.e., all of the irreversible heat goes to electrons.
At $\betai \sim \beta_{\rm i, max}$, the fitting function returns $q_{u \rm e, irr, fit}\approx0.5$, independent of $\bg$, $\sigma_w$, and $\teti$. 
 For $\bg$ in the range $0$--$6$, the fitting function in \eq{fit} is plotted in Figs. \ref{fig:que_irr_bth} and \ref{fig:que_irr_others} as dotted lines, showing that it matches well the trends obtained from the simulations.

\begin{figure*}
		\centering
		\includegraphics[width=\textwidth,clip,trim=3cm 0cm 1cm 0.0cm]{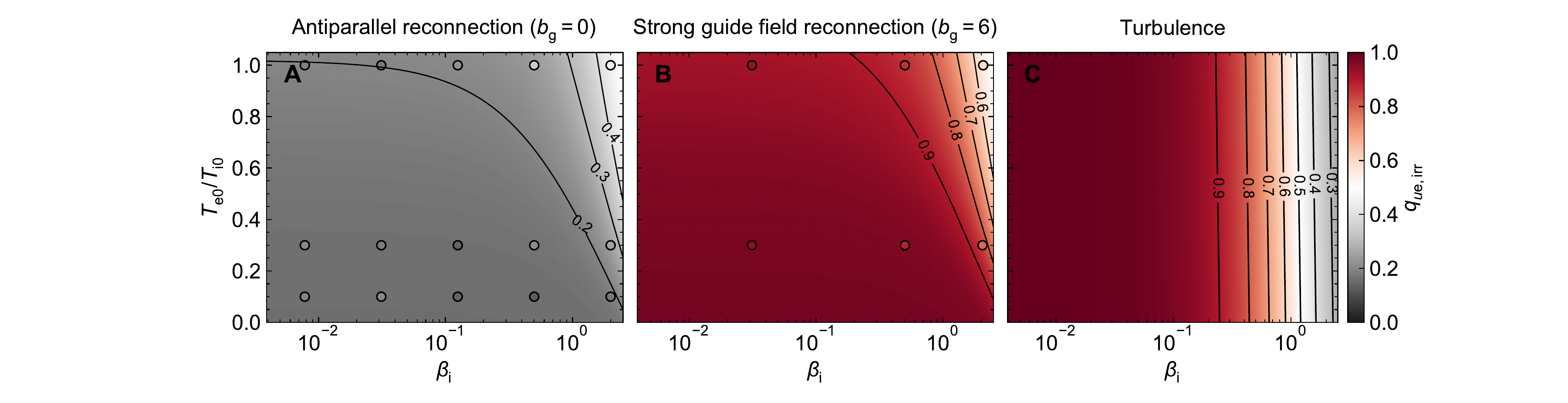} \\
		\caption{Comparison of electron irreversible heating efficiency, $\queirr$ (defined in \eq{queirrev}) for (A) antiparallel reconnection ($\bg=0$), (B) strong guide field reconnection ($\bg=6$), and (C) the turbulent heating prescription of \citet{Kawazura2018} (see Eq. 2 therein), in the $\betai$-$\teti$ parameter space.  Circles in panels A and B show parameters probed directly by the simulations discussed in \sect{others}, colors in panels A and B  employ the fitting function in \eq{fit}.  \label{fig:turb_compare} \\} 
\end{figure*}

Predictions of the reconnection-mediated heating model presented here differ from those of heating via a Landau-damped turbulent cascade \citep{Howes2010,Kawazura2018, Zhdankin2018}.  In \fig{turb_compare} we show a comparison between reconnection-based heating (\eq{fit}) for the antiparallel ($\bg=0$, panel A) and strong guide field ($\bg=6$, panel B) cases, and the turbulence-based heating prescription of \citet{Kawazura2018} (panel C), over the range of plasma conditions we have investigated.  First, one notices that turbulence-based heating is much more similar to heating via reconnection in the strong guide field limit, rather than in the antiparallel case. In fact, for the latter (in contrast to the first two), protons are heated much more than electrons at low $\betai$.
However, some differences persist even between turbulent heating and heating via strong guide field reconnection.  In fact, the turbulence-based heating model is nearly insensitive to the initial temperature ratio $\teti$, whereas for  guide field reconnection, an increase in $\teti$ decreases $\beta_{\rm i, max}$ (see \eq{betaimax}), which in turn decreases the value of $\betai$ at which electrons and protons achieve equipartition, i.e., $\queirr\sim 0.5$. More generally, relativistic effects leave a unique fingerprint in our results at $\betai\sim \beta_{\rm i, max}$,\footnote{We remark that $\beta_{\rm i, max}$ may be much larger or much smaller than unity, depending primarily on $\sigma_w$.} where both species start as relativistically hot, and in the limit $\sigma_w\gg1$. In either case, protons and electrons receive equal amount of the dissipated energy, i.e., $\queirr\sim 0.5$, regardless of the guide field strength.

\subsection{Temperature Anisotropy} \label{sec:anisotropy}
Guide field reconnection can result in highly anisotropic electron distribution functions at late times \citep{Dahlin14, Numata15}.  Yet, to determine the dimensionless internal energy per particle in the fluid rest frame, we have assumed an isotropic stress-energy tensor at every location in the upstream and in the downstream.  \eq{nuint} relies on this assumption.  In addition, we have implicitly assumed isotropy in our prediction for the amount of adiabatic heating.

\begin{figure}
		\centering
		\includegraphics[width=0.45\textwidth,clip,trim=0cm 0cm 0cm 0.0cm]{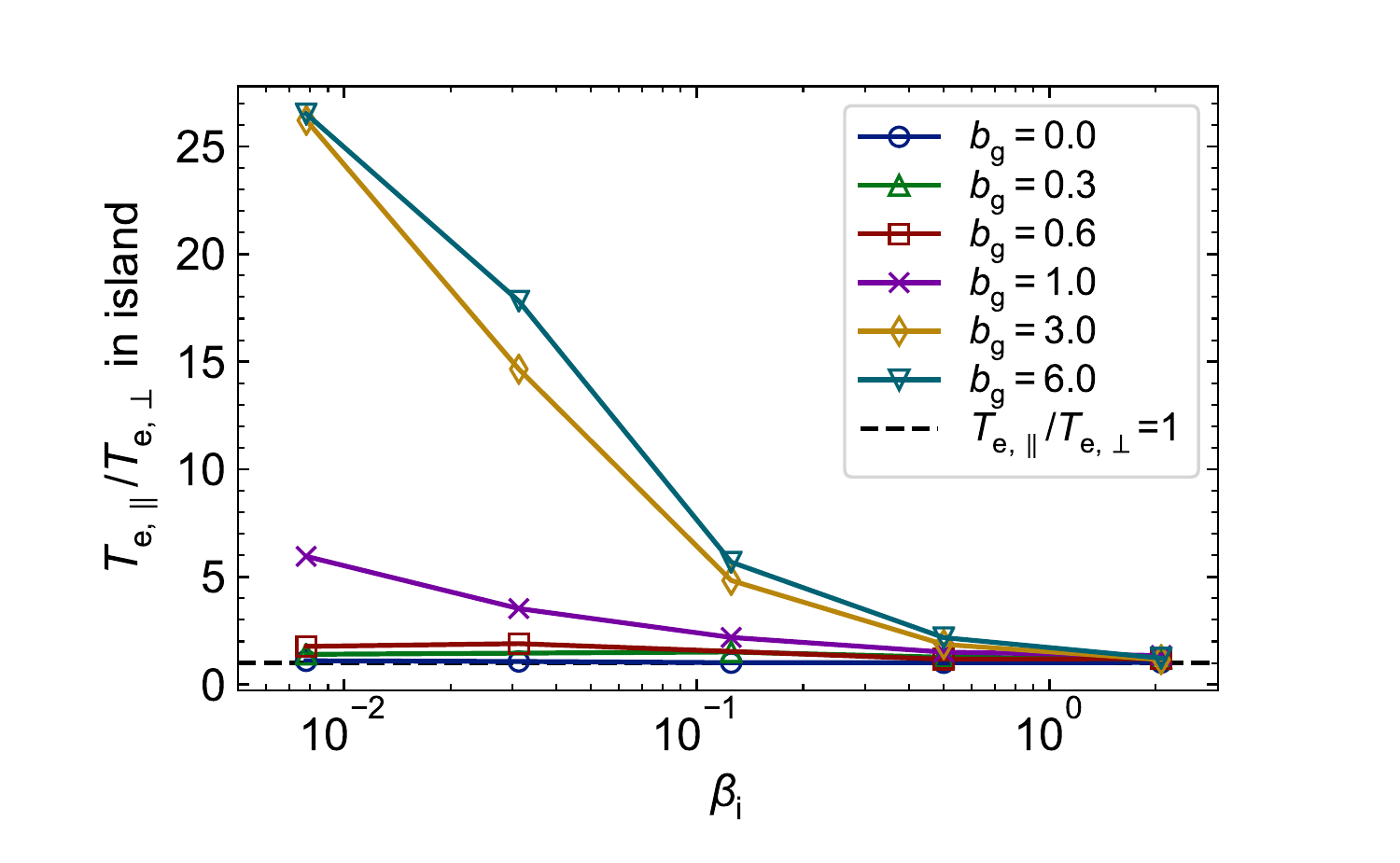} \\
		\caption{Ratio of electron parallel-to-perpendicular temperature in the island region.  For reference, the black dashed line indicates temperature isotropy, $T_{\rm e,\parallel}/T_{\rm e,\perp}=1$.  Parallel ($\parallel$) and perpendicular ($\perp$) are in reference to the direction of the local magnetic field. The simulations shown here are similar to those listed in \tab{params}, but cover $\betai$ in the narrower range $8 \times 10^{-3}$--$2$.\label{fig:aniso_bth} \\} 
\end{figure}

To assess whether isotropy is a reasonable assumption, we show in \fig{aniso_bth} the electron temperature anisotropy $T_{\rm e,\parallel}/T_{\rm e,\perp}$ in the island ($\parallel$ and $\perp$ refer to orientations relative to the local magnetic field); the simulations here are similar to the production runs listed in \tab{params}, but cover $\betai$ more densely in the range $8 \times 10^{-3}$ up to $2$.  

For weak guide fields ($\bg \lesssim 0.3$), the electron temperature is isotropic, $T_{\rm e,\parallel}/T_{\rm e,\perp}\approx 1$ (see \citetalias{rsn17}).  
For $\bg \gtrsim 0.6$ and  $\betai \lesssim 0.5$, we find substantial anisotropy, with temperature ratios in the range $T_{\rm e,\parallel}/T_{\rm e,\perp}\approx 2$--$27$.  In these cases, isotropy is certainly not a valid assumption. As discussed above, this will affect our inferred internal energy (since, in principle, \eq{nuint} cannot be employed) and the predicted degree of adiabatic heating. As regard to the internal energy, in a few cases we have calculated all the components of the stress energy tensor in the simulation frame. By transforming into the comoving frame, we do not need to rely on any assumption of isotropy. In general, we have found the inferred internal energies differ from \eq{nuint} only at the $\sim 10\%$ level.
 
As regard to adiabatic heating, we have discussed in  \sect{gfheating} that compressive heating is suppressed by strong guide fields, as well as at low values of $\betai$.  Therefore, in the majority of cases that show substantial temperature anisotropy, adiabatic heating constitutes a negligible fraction of the total heating, so the degree of anisotropy has only a negligible effect on the inferred irreversible heating.  A notable exception here is the run with $\betai \approx 0.1$ and $\bg=1$, for which compressive heating accounts for about 33\% of the total heating, and the measured anisotropy in the island is non-negligible, $T_{\rm e,\parallel}/T_{\rm e,\perp}\approx 2$; of all our simulations, this one has the greatest systematic uncertainty on the compressive heating, and consequently on the inferred irreversible heating.

\subsection{Mechanisms of Electron Heating in Guide Field Reconnection}\label{sec:mechanism}
The orbit of a charged particle in electromagnetic fields may be approximated as the superposition of two motions: fast circular motion about a point, the \textit{guiding center}, and a slow drift of the guiding center itself.  This approximation is valid when the particle's gyroperiod is short compared to the timescale of variation of the fields, and also when the particle's Larmor radius is small compared to the field gradient length scale.  When valid, the guiding center approximation can provide valuable insight into the mechanisms responsible for particle energization (see e.g., \citet{Dahlin14, Sironi2015, Wang2016}).  In this section, we use the guiding center approximation to investigate the mechanisms of electron heating for $\betai \sim 0.01$, as a function of the guide field strength.  Details of the guiding center decomposition are discussed in \app{appb}.

\begin{figure}
		\centering
		\includegraphics[width=0.45\textwidth,clip,trim=2cm 0.5cm 3cm 0.0cm]{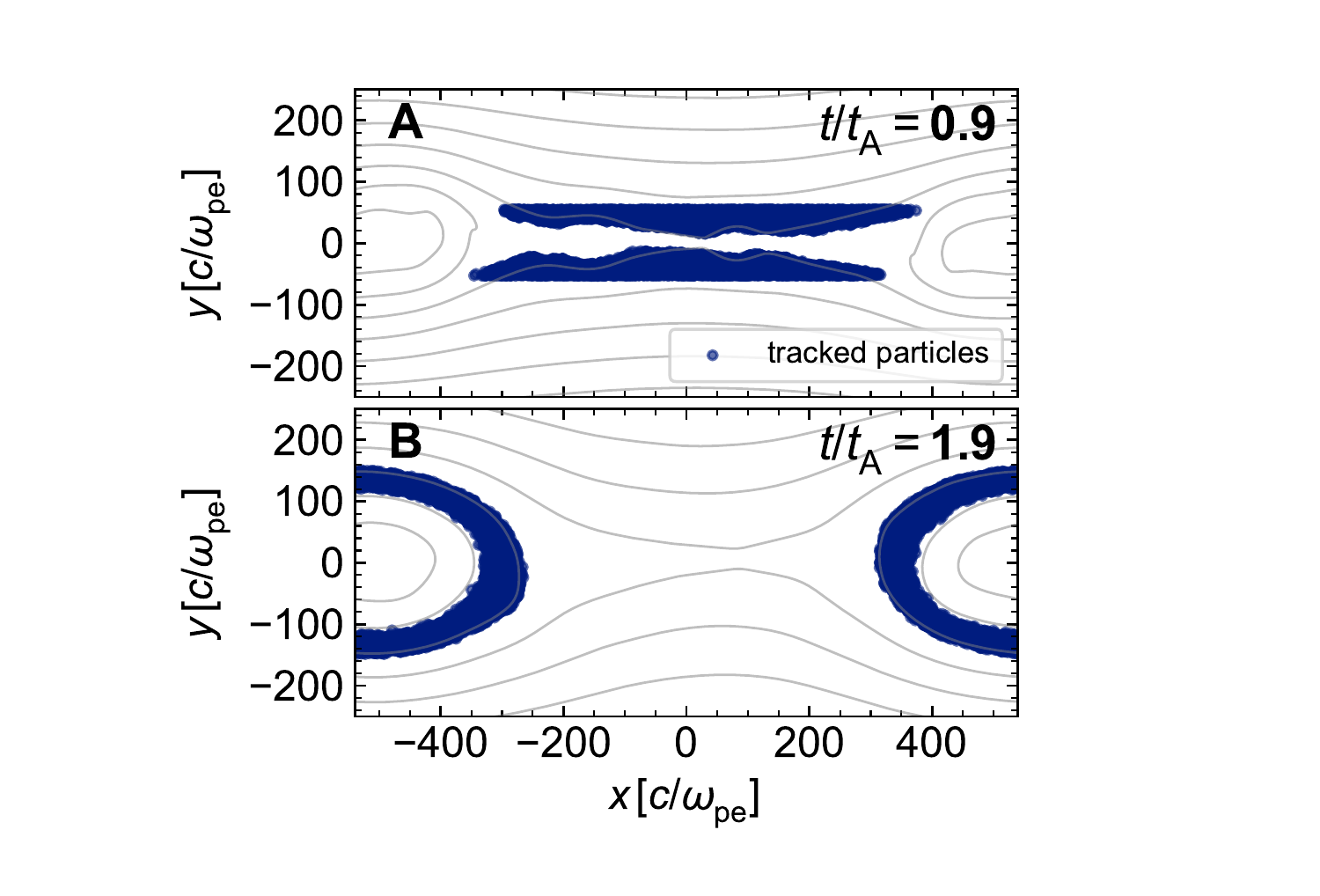} \\
		\caption{Representative (A) initial and (B) final locations of electrons in simulations used for the guiding center analysis outlined in \sect{mechanism}.  For the simulation shown here, $\bg=0.1$, $\betai\approx 7.8 \times 10^{-3}$, $\sigmaw =0.1$, $\mime=1836$, and $\teti=1$.  In this case, a sample of about $1.5\times 10^4$ particles is tracked as they propagate from the upstream to the island region.  \label{fig:prtl_bth_init_final}} 
\end{figure}

We track ${\sim}10^{4}$ electrons starting initially in the upstream region (see \fig{prtl_bth_init_final}, panel A), and compute the contributions $\Delta\varepsilon_{E_{\parallel}}$ and $\Delta \varepsilon_{\rm curv}$, which correspond to energy changes due to the parallel electric field and curvature drift, respectively.  For clarity we focus in our discussion only on the $E$-parallel and curvature drift terms, which tend to dominate for the cases we investigate here (we have directly verified this, and it agrees with findings of \citet{Dahlin14} for nonrelativistic reconnection).  While the simulation timestep is $\Delta t \approx 0.1 \ompinv$, the time interval we use here for outputs of the field and electron properties for the guiding center analysis is around $\Delta t_{\rm out} \approx 3 \ompinv$.  To ensure that this time resolution is sufficient for a guiding center reconstruction, we compare the actual evolution of the electron energy (computed on the fly by the simulation) to the value calculated from the downsampled field and particle information.
For the time range over which we track particles (${{\sim}3700 \ompinv} \approx 1 \ta$ for these simulations), the energy gain computed from downsampled field and particle  information shows excellent agreement with the actual value evolved at the time resolution of the simulation.

To study electron heating via the guiding center theory, we use four simulations for which $\betai=7.8 \times 10^{-3}$ and $\bg \in \{0.1, 0.3, 0.6, 1 \}$.  Here, we use a smaller box size, $L_{x} \approx 1080 \comp$ (the domain size dependence of our results is discussed in \app{appa}).  Apart from the domain size, the parameters are the same as in the main guide field simulations (i.e., $\mime=1836$, $\sigmaw=0.1$, $\teti=1$, $\comp=4$ cells, $N_{\rm ppc} = 16$).  The heating fractions extracted from these simulations are roughly the same as in the production runsof \tab{params}.

Electrons are tracked from $t/\ta \approx 0.9$ to $1.9$ (equivalently, $\tomp \approx 3330$ to $7030$).  The tracked particles are selected at the initial time to lie in the upstream region, within roughly $\pm 50 \comp$ of $y=0$ (see \fig{prtl_bth_init_final}, panel A; grey contours show magnetic field lines).  The selected electrons are tracked for ${{\sim}3700 \ompinv} \approx 1 \ta$, at which point they typically reside in the island region (panel B).

\begin{figure*}
		\centering
		\includegraphics[width=\textwidth,clip,trim=0cm 0cm 1cm 0.0cm]{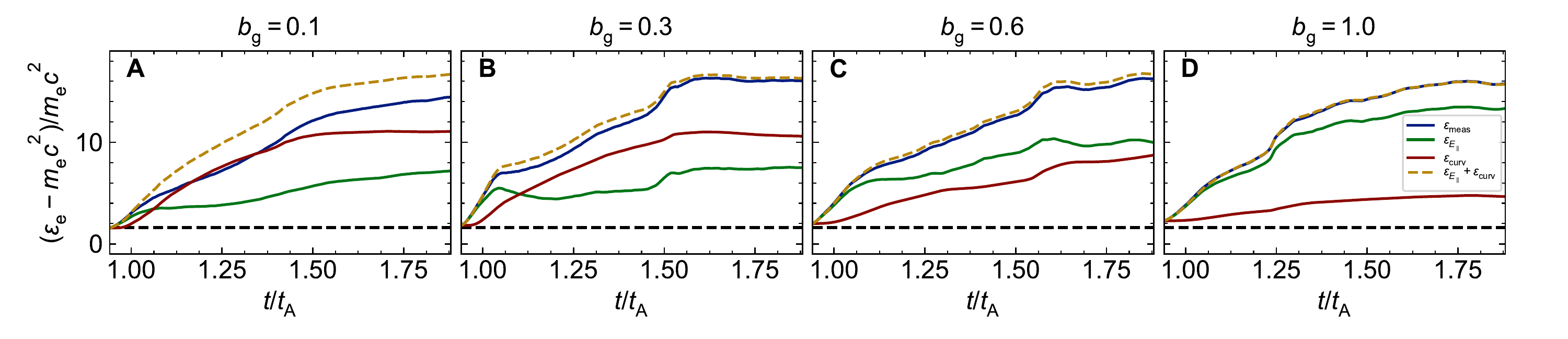} \\
		\caption{Average energy gain per electron for the population of tracked particles (see \fig{prtl_bth_init_final}), for guide field strengths $\bg$ in the range 0.1--1, shown in panels A--D. The measured change in energy (blue) is compared to energization due to the parallel electric field (green) and curvature drift (red) terms, as well as their sum (dashed yellow).  The initial dimensionless electron internal energy in the upstream, $\upsilon_{\rm e0} \approx 1.6$, is shown by the horizontal dashed black line.\label{fig:prtl_bth}}
\end{figure*}

\fig{prtl_bth} shows the time evolution of electron energy gains, for guide fields $\bg =0.1, 0.3, 0.6$, and $1$ in panels A, B, C, and D, respectively (the strength of the guide field increases from left to right).  The energy gain is presented in dimensionless form with rest mass subtracted, i.e., $(\varepsilon_{\rm e} - m_{\rm e} c^2)/m_{\rm e} c^2 \cong \upsilon_{\rm e}$.\footnote{This is not an equality because $\varepsilon_{\rm e} - m_{\rm e} c^2$ includes bulk kinetic energy, in addition to internal energy. However, in the primary island the latter greatly dominates over the former. }  In each panel, the blue line corresponds to the electron energy gain measured directly in the simulations.
 The $E$-parallel and curvature terms are shown in green and red, respectively, and the yellow dashed curve is their sum.  The good agreement between dashed yellow and blue lines is an indication that our output time resolution is adequate for the guiding center reconstruction.  The black dashed line shows the specific internal energy in the far upstream ($\upsilon_{\rm e0} \approx 1.6$), which matches well the starting point of the curves. 

For weak guide fields, $\bg \lesssim 0.6$, the energy gains due to $E$-parallel and curvature terms are comparable, consistent with the findings of \citet{Dahlin14}.  For strong guide fields, energization due to the parallel electric field dominates; in this case, the magnetic field in the current sheet is approximately straight (since it is dominated by the out-of-plane field), so heating due to the curvature term is negligible.  Though the mechanisms responsible for energization of electrons differ for weak and strong guide fields, the overall energy gain is about the same in all cases, $(\varepsilon_{\rm e} - m_{\rm e} c^2)/m_{\rm e} c^2 \cong \upsilon_{\rm e} \approx 14.5$ (see also \fig{dindout_bth}, panel A).  The temporal evolution of electron heating (both the total heating, as well as $E$-parallel and curvature contributions) saturates at late times, when most of the particles reside in the primary island. 

\begin{figure*}
		\centering
		\includegraphics[width=\textwidth,clip,trim=0cm 0cm 1.6cm 0cm]{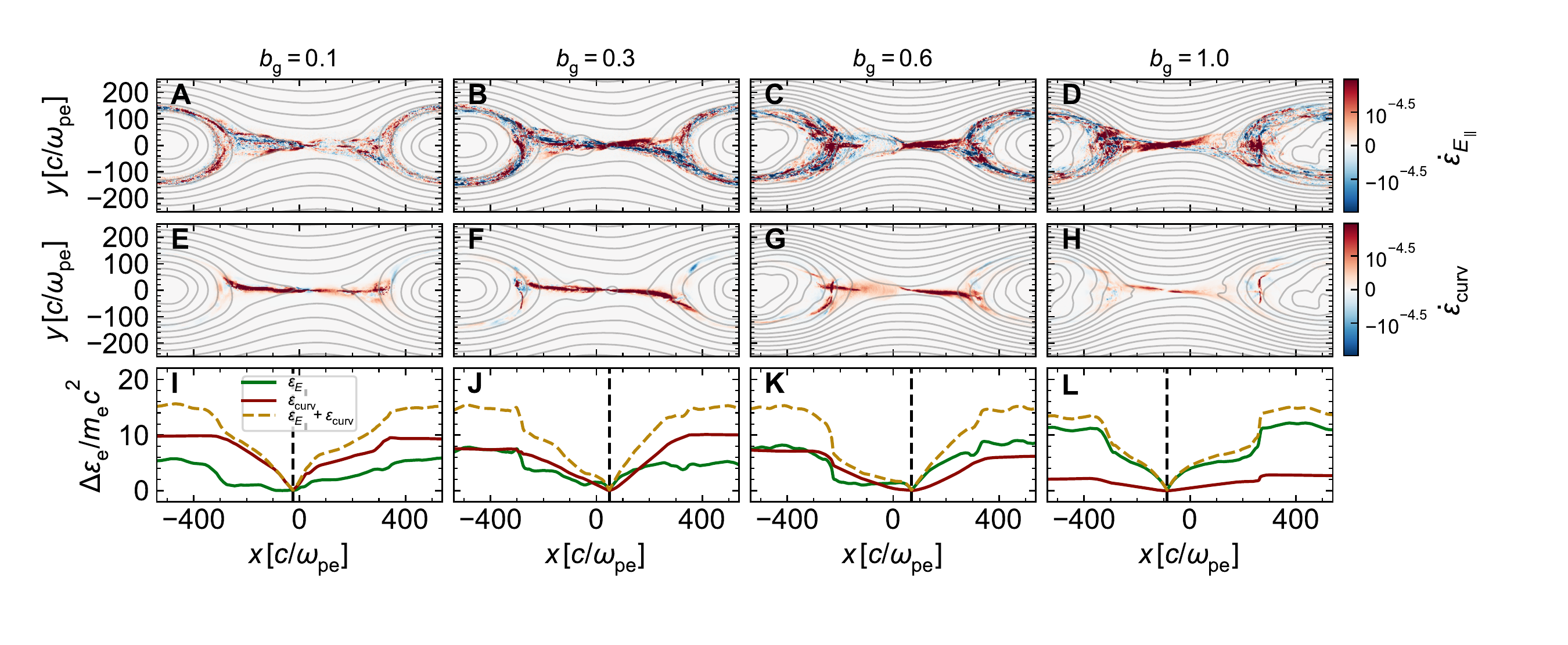} \\
		\caption{2D spatial dependence of energy changes due to (A--D) $E$-parallel and (E--H) curvature drift terms in the guiding center approximation, and (I--L) 1D profiles of $E$-parallel heating (green), curvature drift heating (red), and their sum (dashed yellow).  From left to right, the guide field increases from 0.1 to 1.  The first and second rows
show the per-particle average power deposited from $t/\ta=0.9$ up to $1.9$ ($\tomp\approx 3330$ to $7030$ in these simulations; the domain size $L_{x}\approx1080 \comp$ used here is a factor of two smaller than for the runs in \tab{params}, a choice that is justified in \app{appa}).  For reference, magnetic field lines at $t/\ta=1.9$ are shown in grey, but note that they are not static from $t/\ta=0.9$ to $1.9$.  In the third row, 1D profiles are computed from 2D profiles by summing first along $y$, then summing cumulatively along $x$, starting from the vertical dashed black line (which represents roughly the location of the central X-point) and proceeding outward along ${\pm}x$.\label{fig:prtl_bth_spatial} \\} 
\end{figure*}

\fig{prtl_bth_spatial} shows the 2D spatial distribution of power associated with the $E$-parallel (panels A--D) and curvature (panels E--H) energization terms.  For every tracked electron at each time, we deposit the corresponding $E$-parallel and curvature powers at the location where the particle instantaneously resides (power is deposited into spatial bins of length and width equal to $2 \comp$; note that the colorbar range in \fig{prtl_bth_spatial} depends on this binning, so the units are arbitrary), and then we average over the number of tracked electrons.  Grey lines show the magnetic field lines at $t/\ta\approx 1.9$, for reference.  For weak guide fields ($\bg \lesssim 0.3$), energization due to the parallel electric field is patchy (\citet{Dahlin14}), with heating spread over the exhaust region as well as the island.  On average, there is a net energy gain, however, parallel electric fields can also locally cool the electrons (blue patches in panel A).   Heating due to the curvature drift is localized predominantly along the walls of the exhaust, in particular on the upper left and lower right (as in panel B, for $\bg=0.3$), where outflowing electrons tend to get focused (see also \fig{time_evol2}, panel F).  

As the strength of the guide field increases, the relative importance of the curvature drift energization decreases (panels G--H), and the $E$-parallel heating becomes dominant (panels C--D).  A substantial amount of heating due to the parallel electric field is localized in the exhaust region, however energization continues into the island.  

While the guiding center formalism makes no distinction between adiabatic and irreversible heating, we can infer based on our results for low $\betai$ guide field reconnection (see \fig{mu_bth}, first row) that the $E$-parallel and curvature drift terms in this case (having $\betai\sim0.01$) contribute predominantly to the irreversible heating of electrons.  Since compressive heating is negligible at $\betai\approx 8 \times 10^{-3}$ (see \fig{mu_bth}, panel B; also, \eq{compapprox}), irreversible heating in the low $\betai$ regime represents the main contribution to total electron heating.  It follows that, in this low $\betai$ regime, the guiding center decomposition assesses contributions to irreversible heating.

To clarify the spatial dependence of $E$-parallel and curvature drift heating, we show in the last row of \fig{prtl_bth_spatial} (panels I--L) the 1D cumulative sum along $\pm x$ (as in \citet{Dahlin14}), starting from the vertical dashed line, of the $E$-parallel (solid green) and curvature (solid red) energization rates displayed in the first and second rows; their sum is shown by the dashed yellow line.  This shows that heating continues throughout the exhaust region, and at the interface between the outflow and the primary island. Little additional heating happens inside the primary island.

\section{Summary and discussion} \label{sec:conclusion}
By means of fully-kinetic large-scale 2D PIC simulations, we have investigated guide field reconnection in the transrelativistic regime most relevant to black hole coronae and hot accretion flows.  In particular, we have focused on the fundamental question of electron and proton heating via reconnection, differentiating between adiabatic-compressive and irreversible components. All our simulations employ the realistic mass ratio, $\mime=1836$.  

We find that the energy partition between electrons and protons can vary substantially depending on the strength of the guide field.  For a strong guide field $\bg=B_{\rm g}/B_0 \sim 6$ and low proton beta $\betai \lesssim 0.5$, around $10\%$ of the free magnetic energy per particle is converted to irreversible electron heating (regardless of $\betai$), whereas the efficiency of irreversible proton heating is much smaller, of order ${\sim}1\%$ (these values refer to  our fiducial magnetization $\sigmaw=0.1$ and temperature ratio $\teti=1$).  It follows that the energy partition at high guide fields differs drastically from the antiparallel limit ($\bg=0$), in which electrons receive only ${\sim}6\%$ of the free magnetic energy per particle, and proton irreversible heating is around four times as much, ${\sim}24\%$ \citepalias{rsn17}.

While the energy partition between electrons and protons changes drastically 
with the guide field strength at low $\betai$, at $\betai \sim 2$ ($\approx \beta_{\rm i, max}$, for $\sigmaw=0.1$ and $\teti=1$), the irreversible heating of electrons and protons is in approximate equipartition, regardless of the guide field strength. That is, as $\betai \to \beta_{\rm i, max}$ (when both species start relativistically hot), electrons and protons each receive roughly the same amount of energy, ${\sim}10\%$ of the free magnetic energy per particle in the upstream.

In addition to a comprehensive investigation of the guide field dependence of electron and proton energy partition for our fiducial cases with $\sigmaw=0.1$ and $\teti=1$, we study several cases with larger magnetization, $\sigmaw=1$, and smaller temperature ratios, $\teti=0.1, 0.3$.  Motivated by our extensive exploration of the parameter space (\tab{params} and (\tab{params2}),  we provide a fitting function (\eq{fit}), which captures the approximate dependence of electron irreversible heating efficiency on $\betai, \bg, \sigmaw$, and $\teti$.  This fitting function can be used for sub-grid models of low-luminosity accretion flows such as \sgra\ at the Galactic Center.

As we have said, for strong guide fields and low $\betai$, electrons receive most of the irreversible heat that is transferred to the particles. This is similar to recent findings of electron and proton heating in magnetized turbulence (see Fig.~\ref{fig:turb_compare}, which compares with  \citealt{Kawazura2018}), suggesting  a fundamental connection between reconnection and turbulence, as indeed supported by recent theoretical works
\citep{Boldyrev2017, Loureiro2017, Mallet2017, Comisso2018, Shay2018}. Still, some key differences between our reconnection-based heating prescription and turbulence-based heating prescriptions \citep{Howes2010,Kawazura2018} persist: for $\betai \sim \beta_{\rm i, max}$, when both electrons and protons start relativistic, reconnection leads to equipartition between the two species independently of the guide field strength, whereas for the prescription of, e.g., \citet{Kawazura2018}, protons receive the majority of the irreversible heating at high $\betai$.  Also, for reconnection-based heating, the transition to equipartition happens not at $\betai \sim 1$, but generally at $\betai \sim \beta_{\rm i, max}$, which can differ from unity if $\sigmaw \ll 1$ or $\sigmaw \gg 1$.


We have also used a guiding center analysis to study the mechanisms responsible for electron heating as a function of the guide field strength, for a representative low-$\betai$ case with $\betai \sim 0.01$.  The $E$-parallel and curvature drift terms dominate the energy change of electrons, and their relative importance shifts depending on the strength of the guide field; for weak to moderate guide fields, $0.1 \lesssim \bg \lesssim 0.6$, the energy gains due to $E$-parallel and curvature drift are comparable, but for a strong guide field, $\bg\gtrsim1$, electron energization is dominated by $E$-parallel heating.  Though the mechanisms of electron heating differ depending on the strength of the guide field, the net increase in electron energy remains about the same.  

We conclude by remarking on some simplifying assumptions of the present work, as well as discussing future lines of inquiry.  First, in our investigation of guide field reconnection, we have focused primarily on one value of the magnetization, $\sigmaw=0.1$, and equal temperature ratios in the upstream, $\teti=1$,\footnote{The fitting function \eq{fit}, however, also incorporates results from additional simulations with $\sigmaw=1$ and $\teti=0.1, 0.3$, for both low and high guide field regimes.} to simplify the parameter space investigation.  The dependence of energy partition via reconnection on guide field strength for other values of the magnetization remains under-explored, especially for the low $\betai$ regime, where we find that the proton irreversible heating efficiency depends strongly on the guide field strength.  Similarly, the effect of the upstream temperature ratio $\teti$ in guide field reconnection is under-explored.  

A second simplification is that we have used 2D simulations, which may differ from 3D as regard to particle heating.  In 3D reconnection, in place of magnetic islands, twisted tubes of magnetic flux will develop; to understand the differences as regard to heating, a comparison between 2D and 3D transrelativistic reconnection will be important, especially in the low-$\betai$ regime, where secondary magnetic islands are copiously generated.  

Finally, in our guiding center analysis, we have focused on electron heating in the low $\betai$ regime, where the assumption that the magnetic field varies negligibly over the electron radius of gyration is easily satisfied.  At high $\betai$, this assumption is less robust, and the guiding center theory may  be not applicable.  
Additional theoretical work will be necessary to provide insight into the physics of electron and proton heating in these regimes.

\section*{Acknowledgements}
This work is supported in part by NASA via the TCAN award grant NNX14AB47G and by the Black Hole Initiative at Harvard University, which is supported by a grant from the Templeton Foundation.  LS acknowledges support from DoE DE-SC0016542, NASA Fermi NNX-16AR75G, NASA ATP NNX-17AG21G, NSF ACI1657507, and NSF AST-1716567.  The simulations were performed on Habanero at Columbia, on the BHI cluster at the Black Hole Initiative, and on NASA High-End Computing (HEC) resources.  This research also used resources of the National Energy Research Scientific Computing Center, a DOE Office of Science User Facility supported by the Office of Science of the U.S. Department of Energy under Contract No. DE-AC02-05CH11231.
\appendix

\section{Appendix A:  Convergence of irreversible heating fractions with respect to domain size \texorpdfstring{$L_{\MakeLowercase{x}}$}{}}\label{sec:appa}
\begin{figure*}
		\centering
		\includegraphics[width=\textwidth,clip,trim=0.4cm 0cm 0.8cm 0.0cm]{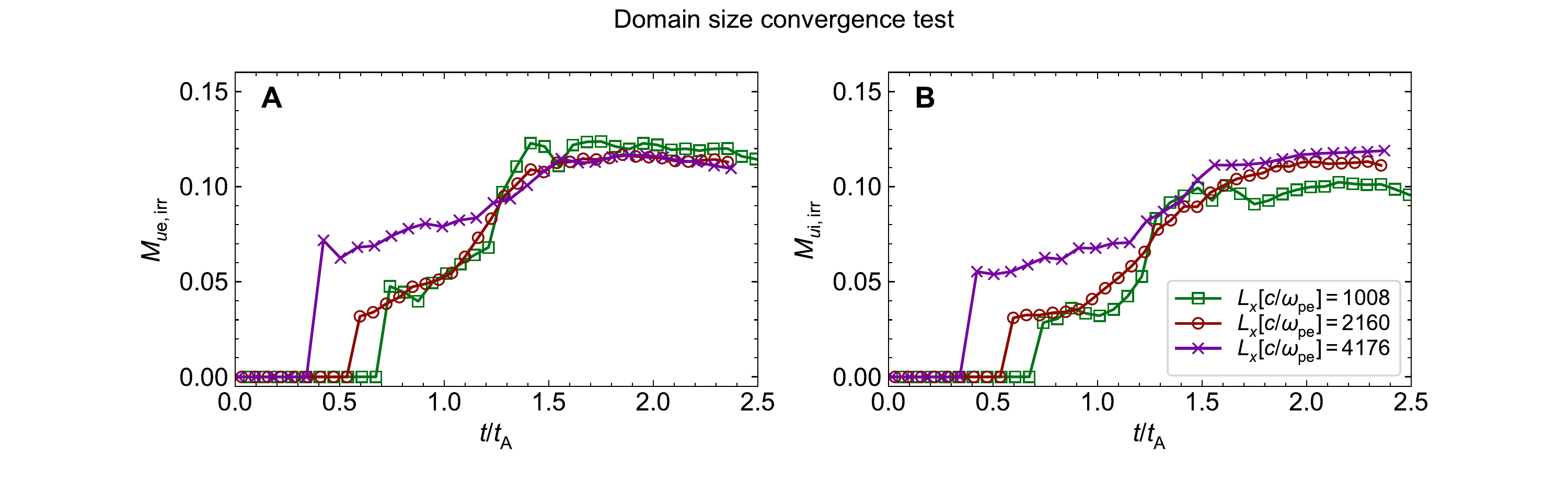} \\
		\caption{Test of convergence for (A) electron and (B) proton irreversible heating fractions, with respect to domain size $L_x$.  We show measured irreversible heating fractions from simulations with $L_x = 1080 \comp$, $2160 \comp$, and $4176 \comp$.  For these simulations, $\bg=1$, $\mime=1836$, $\sigmaw=0.1$, $\betai=0.125$, and $\teti=1$.
		\label{fig:boxsize_conv} \\} 
\end{figure*}

The main focus of this paper is the irreversible heating of electrons and protons.  In principle, the measured values may depend on the domain size $L_x$ if, for example, the computational box is so small that the reconnection outflows do not have the chance to reach the asymptotic Alfv\'{e}n limit.  In this case, the bulk energy of the outflows would be artificially suppressed, which could in turn suppress the particle irreversible heating.  In this appendix, we present a set of lower resolution (\!$\comp=2$ instead of $\comp=4$) simulations with varying domain sizes, $L_x = 1080 \comp, 2160 \comp$ (our fiducial choice), and $4176 \comp$, to explore the box size dependence of the electron and proton irreversible heating fractions, $\mueirr$ and $\muiirr$.  For these simulations, $\bg=1$, $\mime=1836$, $\sigmaw=0.1$, $\betai=0.125$, and $\teti=1$.  In \fig{boxsize_conv}, we show the time dependence of electron (panel A) and proton (panel B) irreversible heating fractions for the three simulations with varying $L_x$.  

For box sizes $L_x \gtrsim 2160 \comp$, the electron irreversible heating converges to $\mueirr \approx 0.11$.  With respect to electron irreversible heating, even the smaller box with $L_x \gtrsim 1080 \comp$ differs by only ${\sim}10\%$ compared to the larger boxes.  This justifies the fiducial domain size $L_x=2160 \comp$ that we use to study electron heating in guide field reconnection, and also the choice of $L_x=1080 \comp$ in \sect{mechanism}, where we use the guiding center theory to study electron energization.  The proton irreversible heating depends more strongly on the box size, but still shows reasonable agreement between the fiducial box size ($L_x=2160 \comp$) and larger boxes ($L_x=4176 \comp$).  In contrast, smaller boxes underestimate the proton heating fraction (green line).

\section{Appendix B:  Guiding Center Formalism}\label{sec:appb}
In the guiding center formalism \citep{Northrop1961, Northrop1963a, Northrop1963b}, the energy change of an electron, time-averaged over the gyration period, and to first order in the expansion parameter $m_{\rm e}/e$, is
\begin{align}
\begin{split}
\frac{1}{e}
&\frac{d \varepsilon_{\rm e}}{d t} =  
	\underbrace{\left[-v_{\parallel} E_{\parallel}  + \frac{ \gamma m_{\rm e}}{e} v_{\parallel} \mathbf{u_{E}} \cdot\frac{d \bhat }{dt} \right]}_{\rm parallel} \\
	+&\underbrace{\left[ \frac{1}{\gamma} \frac{\mu}{e} \mathbf{u_{E}} \cdot \Del B + \frac{1}{\gamma} \frac{\mu}{e} \frac{\partial B}{\partial t} +  \frac{\gamma m_{\rm e}}{e} \mathbf{u_{E}} \cdot \frac{d \mathbf{u_{E}}}{dt}  \right] }_{\rm perpendicular} + \bigO{\left(\frac{m_{\rm e}}{e}\right)^2} \label{eq:energygain}
	 \end{split}
\end{align}
where $\varepsilon_{\rm e}=\gamma m_{\rm e} c^2$, $-e$ is the electron charge, $\mu = \gamma^2 v_{\perp}^2 m_{\rm e}/2B$ is the adiabatic moment of the electron, $\bhat=\bvec/B$ is the direction of the local $B$-field, and $\mathbf{u_{E}}=c\evec \times \bvec /B^2$ is the drift velocity; electric and magnetic fields are to be evaluated at the location of the guiding center.  The underbrackets indicate terms that are associated with parallel and perpendicular energy changes.  Several of the terms have direct physical significance, and provide insight into the mechanisms responsible for particle energization, as we discuss below.

The first of the terms labeled `parallel' in \eq{energygain} corresponds to acceleration by the electric field, parallel to the local $B$-field.  The second term contains the well-known `curvature drift', 
\bea \label{eq:curva}
\frac{\gamma m_{\rm e}}{e} v_{\parallel}^2 \mathbf{u_{E}} \cdot (\bhat \cdot \Del) \bhat &= \frac{\gamma m_{\rm e}}{e} c v_{\parallel}^2 \evec \cdot  \frac{\bhat \times \left[(\bhat \cdot \Del) \bhat \right]}{B} \\
&\equiv -\evec \cdot \mathbf{v}_{\rm curv} .
\eea
which describes the Fermi-like acceleration of particles due to the magnetic tension of curved field lines \citep{Drake2006, Drake2010}.  On the second line of \eq{energygain}, the first term expresses energy change due to the `$\Del B$-drift',
\bea \label{eq:gradb}
\frac{1}{\gamma}\frac{\mu}{e} \mathbf{u_{E}} \cdot \Del B &= \frac{1}{\gamma}\frac{\mu}{e} c \, \evec \cdot \frac{\bhat \times \Del B}{B} \\
&\equiv -\evec \cdot \mathbf{v}_{\Del B}.
\eea
The second term in the second line of \eq{energygain} corresponds to the induction effect of a time-varying field due to $\Del \times \evec$ acting about the circle of gyration \citep{Northrop1963b}.  Finally, the third term is related to energy change due to `polarization drift', which is driven by time-variation in the electric field:
\begin{align}
\frac{\gamma m_{\rm e}}{e} \mathbf{u_{E}} \cdot \frac{\partial \mathbf{u_{E}}}{\partial t} &= \frac{\gamma m_{\rm e}}{e} c \, \evec \cdot \frac{\bhat \times \frac{\partial \mathbf{u_{E}}}{\partial t}}{B} \\
&\equiv -\evec \cdot \mathbf{v}_{\rm pol}.
\end{align}
In practice, many terms in the expansion of \eq{energygain} can be ignored, as their contribution to the electron energy gain is negligible.  In \sect{mechanism}, we employ the guiding center analysis, as detailed in \citet{Dahlin14},
to assess the mechanisms responsible for energy gain in guide field reconnection.  Formally, the electromagnetic fields are to be evaluated at the location of the guiding center, however if the electron Larmor radius is sufficiently small, relative to the gradient length scale of the magnetic field, then the measured value at the guiding center is similar to that at the particle location.  For the simulations described in \sect{mechanism}, we find that this is a reasonable assumption.

\bibliographystyle{apj}
\bibliography{guide_field}

\end{document}